\documentclass[sts,preprint]{imsart}
\usepackage{amsthm,amsmath,amssymb,natbib}
\RequirePackage[colorlinks,citecolor=blue,urlcolor=blue]{hyperref}

\makeatletter
\def\maxwidth{ %
  \ifdim\Gin@nat@width>\linewidth
    \linewidth
  \else
    \Gin@nat@width
  \fi
}
\makeatother
\usepackage[]{color}
\definecolor{fgcolor}{rgb}{0.345, 0.345, 0.345}

\usepackage{framed}
\makeatletter
 {\par\unskip\endMakeFramed%
 \at@end@of@kframe}
\makeatother

\definecolor{shadecolor}{rgb}{.97, .97, .97}
\definecolor{messagecolor}{rgb}{0, 0, 0}
\definecolor{warningcolor}{rgb}{1, 0, 1}
\definecolor{errorcolor}{rgb}{1, 0, 0}
\newenvironment{knitrout}{}{} % an empty environment to be redefined in TeX

\usepackage{alltt}
\setattribute{journal}{name}{}
\usepackage{graphicx}
\usepackage{multirow}
\usepackage{lscape}
\usepackage{booktabs}
\usepackage{url}
\usepackage{array}
\usepackage{todonotes}
\usepackage{float}
\usepackage[font=it,size=footnotesize,labelfont=sc,labelsep=period]{caption}
\usepackage{subcaption}
\newcommand{\subfloat}[2][need a sub-caption]{\subcaptionbox{#1}{#2}}
\RequirePackage[colorlinks,citecolor=blue,urlcolor=blue]{hyperref}

\newcommand{\R}{\mathbb{R}}
\newcommand{\Ss}{\mathbb{S}}
\newcommand{\E}{\mathbb{E}} 
\newcommand{\prob}{\mathbb{P}}
\newcommand{\V}{\mathbb{V}}
\newcommand{\cov}{\mathrm{Cov}}
\newcommand{\C}{\mathbf{C}}
\newcommand{\K}{\mathbf{K}}
\newcommand{\bx}{\mathbf{x}}
\newcommand{\by}{\mathbf{y}}
\newcommand{\bu}{\mathbf{u}}
\newcommand{\bs}{\mathbf{s}}
\newcommand{\bo}{\mathbf{0}}
\newcommand{\N}{\mathcal{N}}

\newcommand{\ddui}{\frac{\partial}{\partial u_i}}
\newcommand{\ddth}{\frac{\partial}{\partial \theta_i}}

\newcommand{\Cun}{\C_{\theta}(\bu_n)}
\newcommand{\xkale}{\hat{x}_{\text{\tiny\textsc{KALE}}}}
\newcommand{\xkile}{\hat{x}_{\text{\tiny\textsc{KILE}}}}

\newcommand{\bQ}{\mathbf{Q}}
\newcommand{\bR}{\mathbf{R}}
\newcommand{\bD}{\mathbf{D}}
\newcommand{\ba}{\mathbf{a}}
\newcommand{\bb}{\mathbf{b}}
\newcommand{\bLam}{\boldsymbol{\Lambda}}
\newcommand{\bOm}{\boldsymbol{\Omega}}

\newtheorem{theorem}{Theorem}[section]

\newtheorem{proposition}[theorem]{Proposition}

\IfFileExists{upquote.sty}{\usepackage{upquote}}{}
\begin{document}

\begin{frontmatter}

\title{Gaussian Process Regression with Location Errors}
\runtitle{Gaussian Process Regression with Location Errors}

\begin{aug}
\author{\fnms{Daniel} \snm{Cervone} \thanksref{t1} \ead[label=e1]{dcervone@fas.harvard.edu}}
\and
\author{\fnms{Natesh} \snm{S. Pillai}\thanksref{t1} \ead[label=e2]{pillai@stat.harvard.edu}}
\thankstext{t1}{NSP was partially supported by an ONR grant. DC was partially supported by a research grant from the Harvard University Center for the Environment.}

\affiliation{Harvard University\thanksmark{t1}}
\address{Department of Statistics, Harvard University, Cambridge, MA, USA \printead{e1,e2}.}

%\address{
%\printead{e1}}
\runauthor{D. Cervone and N. Pillai}
\end{aug}

\bibpunct{(}{)}{,}{a}{}{;}

\begin{abstract}
In this paper, we investigate Gaussian process regression models where inputs are subject to measurement error. In spatial statistics, input measurement errors occur when the geographical locations of observed data are not known exactly. Such sources of error are not special cases of ``nugget'' or microscale variation, and require alternative methods for both interpolation and parameter estimation. Gaussian process models do not straightforwardly extend to incorporate input measurement error, and simply ignoring noise in the input space can lead to poor performance for both prediction and parameter inference. We review and extend existing theory on prediction and estimation in the presence of location errors, and show that ignoring location errors may lead to Kriging that is not ``self-efficient''. We also introduce a Markov Chain Monte Carlo (MCMC) approach using the Hybrid Monte Carlo algorithm that obtains optimal (minimum MSE) predictions, and discuss situations that lead to multimodality of the target distribution and/or poor chain mixing. Through simulation study and analysis of global air temperature data, we show that appropriate methods for incorporating location measurement error are essential to valid inference in this regime. 
%Our results We discuss  We show that the usual Kriging estimators ignoring the location errors may not be ``self-efficient". We discuss both likelihood and Bayesian perspectives and present extensive simulation studies showing the relative performances and  coverage properties of various estimation procedures. Bayesian model fitting is done using Hybrid Monte Carlo algorithm. We show that the presence of location errors can lead to multimodal posterior distributions and consequently affect the mixing of the HMC sampler. We present an analysis of northern hemisphere temperature data from the summer of 2011 and demonstrate that appropriate methods for incorporating location measurement error are essential to valid inference in this regime.
\end{abstract}

%\begin{keyword}[class=MSC]
%\kwd[Primary ]{}
%\kwd{}
%\kwd[; secondary ]{}
%\end{keyword}


\end{frontmatter}

%A. Intro
% - define GPs
% - present location error model
%B. Kriging (KALE)
% 1. moments
% 2. interval estimates
% 3. parameter estimation (pseudolikelihood)
% 4. advantages over KILE (): admissibility, self-efficiency, of kriging; consistency and effieincy of parameter estimation.
%C. Monte Carlo inference
% 1. requires distributional assumption; gaussian makes sense.
% 2. information in location error, mutual information in location error and prediction
% 3. known parameters: perfect conditional coverage, dominates KALE
% 4. unknown parameters: obviously extends with hmc, choices of prior
% 5. computational/convergence issues (reducing parameter space)
%D. Simulation study
%E. Real Data example
%F. Extensions to large data sets
%G. Conclusion

\section{Introduction}
Gaussian process models assume an output variable of interest varies smoothly over an input space (\textit{e.g.}, percipitation totals across geographical coordinates, crop yield across factor levels of an experimental design). Such models appear frequently in areas as diverse as climate science [\citet{mardia1993spatial}], epidemiology [\citet{lawson1994using}], and black-box problems such as computer experiments, and Bayesian optimization [\citet{sacks1989design, srinivas2009gaussian}]. See \cite{stein1999interpolation, cressie1993statistics, baner14} and \cite{gelman2014bayesian} for more detailed treatments.

Noisy spatial input data are common in many applications; for example, geostatistical data is often imprecisely spatially referenced, ``binned'' to the nearest latitude/longitude grid point, or referenced to maps with distorted coordinates [\citet{veregin1999data, barber2006modelling}]. Accounting for measurement error on covariates in the context of regression models is a well studied theme [\citet{carroll2006measurement}]; however, despite their importance in applications, surprisingly little work has been done on interpolation or Gaussian process regression problems in the presence of (spatial) location measurement error. As we show in this paper, Gaussian process models do not straightforwardly extend to incorporate input measurement error, and simply ignoring noise in the input space can lead to poor performance. \par
Previous research on such error sources has mostly focused on demonstrating their existence and quantifying their magnitude [\citet{bonner2003positional, ward2005positional}]. 
For regression problems, \citet{gabrosek2002effect} (and later \citet{Cressie2003}) adjust Kriging equations for the presence of location errors, and \citet{fanshawe2011spatial} further develop research for this regime to include problems where the locations of future observations or predictions are subject to error. Location errors have also been studied in the context of point process data [\citet{zimmerman2006estimating, zimmerman2010spatial}].

Properly accounting for location errors is essential for optimal interpolation and uncertainty quantification, as well precise and efficient parameter estimation when parameters of the covariance function are unknown. Using theoretical results and extensive simulations, our paper provides guidelines on situations when location errors are most impactful for data analysis, and suggestions for incorporating this source of error into inference and prediction. We expand the research in \citet{Cressie2003} on best linear unbiased prediction (Kriging) to include procedures for obtaining interval forecasts and for quantifying the cost of ignoring location errors. We also discuss Markov Chain Monte Carlo (MCMC) methods for optimal (minimum mean squared error (MSE)) predictions, which average over the conditional distribution of (latent) location errors given the observed data. %We discuss some aspects of model fitting using the Hybrid Monte Carlo (HMC) algorithm implemented using the software \texttt{RStan} [\citet{rstan-software:2014}].

Section \ref{sec:notation} establishes notation and describes the basic model with location errors used throughout the paper. In Section \ref{sec:kriging}, we discuss Kriging using the covariance structure of the location-error induced process. Section \ref{sec:hmc} considers MCMC methods for obtaining minimum MSE predictions, and thus improving upon Kriging. We compare these methods through simulation study in Section \ref{sec:simul}, and explore an application to interpolating northern hemisphere temperature anomolies in Section \ref{sec:cru}. The proofs of all of the theoretical results are given in an Appendix.
%Section 6 briefly considers the location error issue for large data sets, when traditional Gaussian process methodology is computationally infeasible.

%meaning that appropriate methods can deliver interpolations with lower MSE than  Heuristically speaking, this is because the response at noise-corrupted locations, $x(s_i + u_i)$ might be more correlated with $x(s^*)$ than the response at the intended locations $x(s_i)$. Of course, we do not observe location errors $u_i$, but there is information in the data to infer them, which can ultimately be leveraged to obtain a more faithful forecast of $x(s^*)$. 
\section{The Model}\label{sec:notation}
We will write $\bs_n = (s_1 \: \: s_2 \: \ldots \: s_n)'$ to denote a $n$-vector of locations in the input space $\Ss \subset \R^p$, and $\bx_n = (x(s_1) \: \: x(s_2) \: \ldots \: x(s_n))'$ as the associated vector of observations at $\bs_n$. Similarly, we will denote $\bx^*_k = (x(s^*_1) \: \ldots \: x(s^*_k))'$, or simply $x^* = x(s^*)$ where $\{s^*_i, i=1, \ldots, k\}$ are unobserved locations. %Locations are assumed to be $p$-dimensional, so that $s_i \in \Ss \subset \R^p$ for all $i$. 
The process $x: \Ss \rightarrow \R$ is called a \textit{Gaussian process} if, for any $s_1, \ldots s_n \in \Ss$, $\bx_n = (x(s_1) \: \: x(s_2) \: \ldots \: x(s_n))'$ is jointly Normally distributed. Typically, the form of this joint distribution is specified by a deterministic or parametric mean function (for now, taken without loss of generality to be 0) and a covariance function $c: \Ss^2 \rightarrow \R$, so that
\begin{equation}\label{GPdef}
\begin{pmatrix} x(s_1) \\ \vdots \\ x(s_n) \end{pmatrix}
\sim \N \left( \bo, 
\begin{pmatrix} c(s_1, s_1) & \cdots & c(s_1, s_n) \\ 
\vdots & \ddots & \\ 
c(s_n, s_1) &  & c(s_n, s_n) \end{pmatrix}
\right).
\end{equation}
For $c$ to be a valid covariance function, the covariance matrix in Equation \eqref{GPdef} must be positive semi-definite for all input vectors $\bs_n = (s_1 \: \: s_2 \: \ldots \: s_n)'$.
 %One common choice of covariance function is the Mat\'ern, which has the form

%\todo[inline]{Matern covariance, and exponential, sqexp as limiting case}

Gaussian process regression is primarily used as a method for interpolating (predicting) values $\bx^*_k$ at unobserved points $\bs^*_k = (s^*_1 \: \ldots \: s^*_k)'$ in the input space, given all available observations. Such conditional distributions are easily obtained by exploiting the joint normality of the response $x$ at observed and unobserved locations:
\begin{align}
\bx^*_k | \bx_n & \sim \N \big( \C(\bs^*_k, \bs_n)\C(\bs_n, \bs_n)^{-1} \bx_n, \nonumber \\
& \hspace{1.75cm} \C(\bs^*_k, \bs^*_k) - \C(\bs^*_k, \bs_n)\C(\bs_n, \bs_n)^{-1}\C(\bs_n, \bs^*_k)\big).
\label{GPconditional}
\end{align}
In Equation \eqref{GPconditional}, $\C(\bs_n, \bs_n)$ denotes the covariance matrix of $\bx_n$, $\C(\bs^*_k, \bs_n)$ denotes the $k \times n$ covariance matrix between $\bx^*_k$ and $\bx_n$.

When the locations in the input space $\Ss$ are affected by error, we observe a surrogate process $y: \Ss \rightarrow \R$,
$$y(s_i) = x(s_i + u_i),$$ 
where $s_i$ is a known location in $\Ss$ and $u_i \in \Ss$ is unobserved location error. The problem of Gaussian process regression with location errors addressed in this paper is to predict $x$ at unobserved (exact) locations $x(s^*)$ given observations from the noise-corrupted process $y$.

When $x$ is assumed to be a Gaussian process, there is no nontrivial structure for $u$ that results in $y$ being a Gaussian process. Additionally, it is not possible to write $y$ as a convolution of $x$ and a white noise process as differences between the surfaces $y$ and $x$ will generally be correlated across space, \textit{i.e.,} $\mathrm{Cov} [y(s_1) - x(s_1), y(s_2) - x(s_2)] \neq 0$. Gaussian process regression with location errors therefore cannot be thought of as a classical or Berkson errors-in-variables problem [\cite{carroll2006measurement}].
Interestingly, in some cases, the process $y$ may be more informative for prediction at a new location $x(s^*)$ than the process $x$ is.  Thus, appropriate methods can deliver lower MSE interpolations in a location-error regime than the MSE of the usual methods in an error-free regime.

\section{Kriging the Location Error Induced Process $y$}
\label{sec:kriging}

As shown in \citet{Cressie2003}, the second moment properties of $y$ can be used to perform Kriging (they named this ``Kriging adjusting for location error'' or KALE), noting that measurement errors $u$ induce a new covariance function 
\begin{alignat}{2} 
k(s_1, s_2) &= \cov[y(s_1), y(s_2)] & &= \E[c(s_1 + u_1, s_2 + u_2)]  \text{ for $s_1 \neq s_2$ } \nonumber \\
k(s, s) &= \V[y(s)] & &= \E[c(s + u, s + u)] \nonumber \\
k^*(s, s^*) &= \cov[y(s), x(s^*)] & &= \E[c(s + u, s^*)]. 
\label{cov_y}
\end{alignat}
The expectation here is taken over the input errors $u$, which are assumed to have some joint distribution $g_{\bs_n}$. The following result shows that if $c$ is a valid covariance function, then so is $k$, regardless of the error distribution $g(\cdot)$.
%g_{\bs_n} just arbitrary distribution
\begin{proposition}\label{validcovfn}
Assume for all $n$ and $\bs_n \in \Ss$, ${(u_1, u_2, \ldots, u_n) \sim g_{\bs_n}\in \mathcal{G}}$, where $\mathcal{G}$ is any family of probability measures on $\Ss$. Then $k$ is a valid covariance function if $c$ is.
\end{proposition}
Regardless of the form of $c$, $k$ always exhibits the ``nugget'' effect, or discontinuities in the covariance function [\citet{matheron1962traite}] ${\lim_{s_2 \to s_1} k(s_2, s_1) \neq k(s_1, s_1).}$ In fact, several authors cite location/positional error as a justification for including a nugget term in an arbitrary covariance function $c$ [\cite{cressie1993statistics, stein1999interpolation}], alongside independent measurement error in observing the response, $x(s) + \epsilon$. Location errors, however, cause $k$ to differ from $c$ throughout the spatial domain $\Ss^2$ (this is shown in Figure \ref{fig:c_vs_k}), meaning that while they induce a nugget, a nugget term alone cannot capture the effect of location errors.

\begin{knitrout}
\definecolor{shadecolor}{rgb}{0.969, 0.969, 0.969}\color{fgcolor}\begin{figure}[H]

\includegraphics[width=\maxwidth]{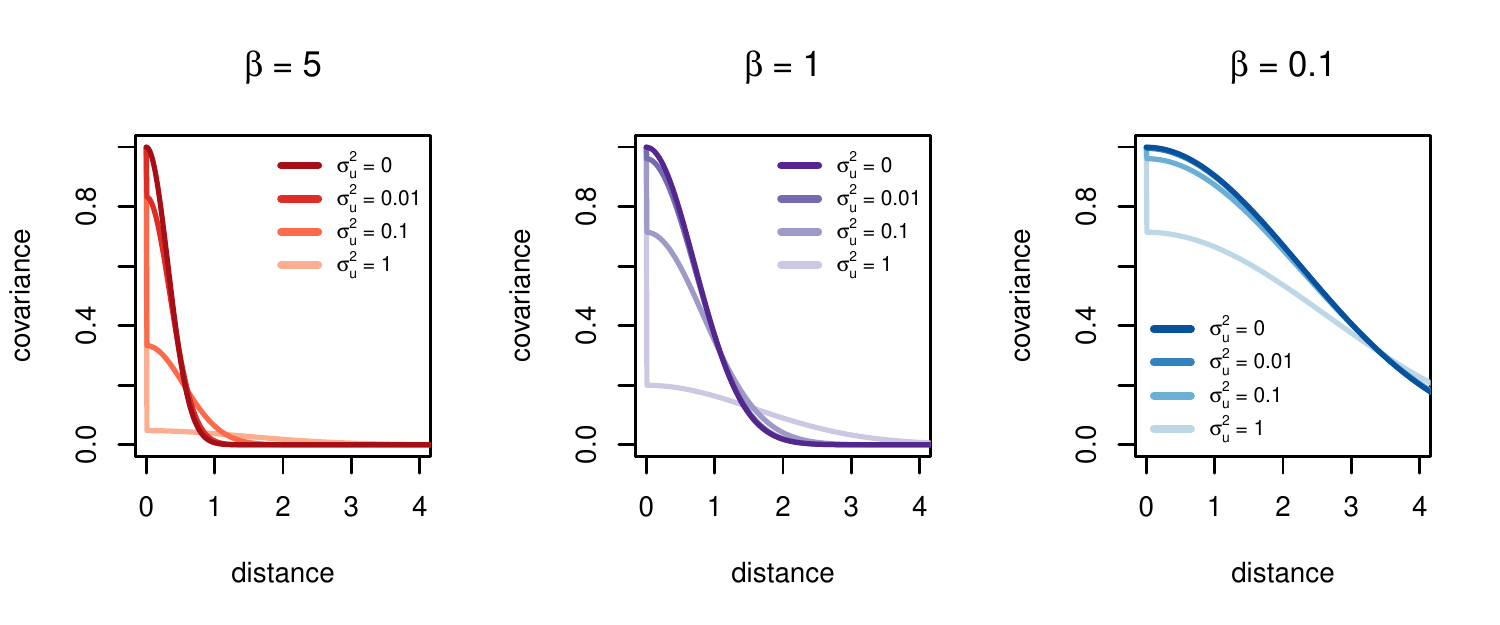} \caption[Comparison of $c$ and $k$ for $\Ss = \R^2$ and $c(s_1, s_2) = \exp(-\beta\|s_1 - s_2\|^2)$, with $u_i \stackrel{iid}{\sim} \N(0, \sigma^2_u\mathbf{I}_2)$]{Comparison of $c$ and $k$ for $\Ss = \R^2$ and $c(s_1, s_2) = \exp(-\beta\|s_1 - s_2\|^2)$, with $u_i \stackrel{iid}{\sim} \N(0, \sigma^2_u\mathbf{I}_2)$. Location errors $\sigma^2_u > 0$ cause $c$ and $k$ to differ as a function of distance, and induce a nugget discontinuity at 0.\label{fig:c_vs_k}}
\end{figure}

\end{knitrout}

Using $k$, we get the Kriging estimator adjusting for location error for $x(s^*)$ at an unobserved location of $x$:
\begin{equation}\label{ykriging}
\xkale(s^*) = \K^*(s^*, \bs_n) \K(\bs_n, \bs_n)^{-1} \by_n.
\end{equation}
In Equation \eqref{ykriging}, $\K$ and $\K^*$ respectively denote the covariance matrices corresponding to the kernels $k$ and $k^*$. The quantity $\xkale(s^*)$ is the best linear unbiased predictor for $x(s^*)$ (in terms of MSE) and has all the usual Kriging properties. When there are no location errors, the Kriging estimator is equivalent to the conditional expectation of $x(s^*)$ given $\bx_n$ (see Equation \eqref{GPconditional}).

In general, the covariance functions $k$ and $k^*$ can be evaluated using Monte Carlo integration by repeatedly sampling $\bu_n$ from $g$. For several common combinations of covariance function and location error models, however, it is possible to arrive at the expressions in Equation \eqref{cov_y} in closed form. In particular, if 
$$c(s_1, s_2) = \tau^2 \exp(-\beta d(s_1, s_2)),$$ 
then we can define a random variable $Z = d(s_1 + u_1, s_2 + u_2)$ and find its moment generating function $M_Z(t)$. If we can evaluate $M_Z(t)$ at $t = -\beta$, then this yields $k(s_1, s_2)$. For instance, for the squared exponential covariance function $d(s_1, s_2) = \|s_1 - s_2\|^2$ and Normal location errors $u \sim \N(0, \sigma^2_u \mathbf{I}_p)$, $Z$ has a scaled noncentral $\chi^2_p$ distribution and 
\begin{align}
k(s_1, s_2) &= \frac{\tau^2}{(1 + 4\beta\sigma^2_u)^{p/2}}\exp \left( - \frac{\beta}{1 + 4 \beta \sigma^2_u} \|s_1 - s_2\|^2 \right) \text{ for } s_1 \neq s_2 \nonumber \\
k(s, s) &= \tau^2
\label{cov_y_sqexp}
\end{align}
with a similar expression for $k^*(s, s^*)$. Thus the covariance function for $y$ is also squared exponential (it is not generally true that $c$ and $k$ will share the same functional form). Note, however, that not all parameters are identifiable---we must know at least one of $(\tau^2, \beta, \sigma_u^2)$ in order to estimate the others. %\todo{cite Cressie miscroscale identifiability distinction?}

Interestingly, it is possible for the KALE to yield lower MSE predictions than those given from an error-free regime, where $\bu_n \equiv 0$ and $x = y$. In other words, $\by_n$ can be more informative than $\bx_n$ for predicting $x(s^*)$. Heuristically, this happens when $\by_n$ is more strongly correlated with $x(s^*)$ than is $\bx_n$. Below we characterize the conditions for observing this phenomenon in a simple model with one observed data point (Figure \ref{fig:c_vs_k} provides an illustration); it seems difficult to generalize  this to larger observed location samples and covariance/error structures.
%\todo[inline]{plot of $c$ versus $k$?}

\begin{proposition} Assume $n=1$, $\|s - s^*\|^2 = \Delta^2$, $c(s, s^*) = \tau^2 \exp(-\beta \Delta^2)$ for all $s, s^* \in \Ss$, and $u \sim \N(0, \sigma^2_u \mathbf{I}_p)$. Without location error ($\sigma^2_u = 0$), the \emph{MSE} in predicting $x(s^*)$ from $x(s)$ is $c_0 = \tau^2(1 - \exp(-2\beta\Delta^2))$. There exists $\sigma_u>0$ such that $\E[(\xkale(s^*) - x(s^*))^2] < c_0$ if and only if $\beta\Delta^2 > p/2$.
\label{prop1} 
\end{proposition}

%Kriging using second moment properties of $y$, the input-error corrupted process, is explored by Cressie and Kornak. The Kriging predictor for the process at new locations, $\bx(\bs^*) = (x(\bs^*_1) \: \: x(\bs^*_2) \: \ldots \: x(\bs^*_k))'$, given a collection of observations $\by(\bs) = (y(\bs_1) \: \: y(\bs_2) \: \ldots \: y(\bs_n))'$ is $\K^*(\bs^*, \bs)\K(\bs, \bs)^{-1}\by(\bs)$. Here, $\K$ is the covariance matrix obtained from the marginal covariance function $k$ \eqref{cov_y} using the observations corresponding to $\bs$, and $\K^*$ is the covariance matrix deriving from a new covariance function $k^*$ for a process where location errors corrupt observations at $\bs$ but not at $\bs^*$:
%\begin{equation}
%\cov[x(\bs^*), y(\bs)] = k^*(\bs^*, \bs) = \E[c(\bs^*, \bs + \bu)]
%\end{equation}
%It is sometimes possible to derive analytic expressions for $k$ and $k^*$, for example, in the case of a squared exponential covariance function and a Gaussian or uniform error distribution $g(\bu)$. Otherwise, it is necessary to compute $\K^*$ and $\K$ by Monte Carlo. Note however, that despite using a Monte Carlo procedure only one matrix inversion is needed for Kriging.

\subsection{Interval predictions}

For many applications of Gaussian process regression, particularly in geostastics and environmental modeling, both point and interval predictions are of interest. However, Kriging, being strictly a moment-based procedure, does not provide uncertainty quantification for predictions other than variance. In a location-error Gaussian process regime, KALE predictions will always be non-Gaussian, thus variance alone is not sufficient to provide distributional or interval predictions.

However, it is relatively straightforward to derive confidence intervals for predictions at unobserved locations $x(s^*)$ given measurements $\by_n$ at locations $\bs_n$. The following proposition provides the exact distribution function (CDF) for prediction errors $x(s^*) - \xkale(s^*)$, which can be inverted to obtain a confidence interval for $x(s^*)$ based on $\xkale(s^*)$.
\begin{proposition}Let
\begin{align*}
V(\bu_n) & = \sigma^2 + \gamma'\C(\bs_n + \bu_n, \bs_n + \bu_n)\gamma - 2\gamma'\C(\bs_n + \bu_n, s^*) \\
\text{where } \: \gamma &= \K(\bs_n, \bs_n)^{-1}\K^*(\bs_n, s^*), \sigma^2 = \V [x(s^*)].
\end{align*}
Then 
\begin{equation}\label{Wcdf}
\prob(x(s^*) - \xkale(s^*) < z) = \E \left[ \Phi \left( \frac{z}{\sqrt{V(\bu_n)}} \right) \right],
\end{equation}
where $\Phi$ is the CDF of the standard Normal distribution. 
\label{coverprop}
\end{proposition}
%Let $W = x(s^*) - \xkale(s^*)$. We can explicitly write the dependence of $W$ on $\bu_n$:
%\begin{equation}\label{Wdist}
%W | \bu_n \sim \N \big( 0, \sigma^2 + \gamma'\C(\bs_n + \bu_n, \bs_n + \bu_n)\gamma - 2\gamma'\C(\bs_n + \bu_n, s^*) \big)
%\end{equation}
%where $\sigma^2 = \V [x(s^*)]$ and $\gamma = \K(\bs_n, \bs_n)^{-1}\K^*(\bs_n, s^*)$; that is, $\xkale(s^*) = \gamma'\by_n$. 
%Thus $\prob(W < z) = \E[\prob(W < z | \bu_n)]$. Letting $V(\bu_n)$ be the variance term in \eqref{Wdist}, we get 
It may be necessary to evaluate Equation \eqref{Wcdf} using Monte Carlo; if so, it is practical to use the same draws of $\bu_n$ when evaluating different quantiles $z$, as this guarantees a Monte Carlo estimate of the distribution function be non-decreasing.

%Thus, for any $\alpha$ we can find values $z_{\text{low}}, z_{\text{hi}}$ such that $\prob(z_{\text{low}} < W < z_{\text{hi}}) = 1 - \alpha$. This probability statement can be ``inverted'' to yield $(\gamma' \by_n + z_{\text{low}}, \gamma' \by_n + z_{\text{hi}})$ as a $1 - \alpha$ confidence interval for $x(s^*)$. 
While intervals based on \eqref{Wcdf} provide exact coverage (modulo Monte Carlo error), such coverage is achieved by averaging over all data, both observed ($\by_n$) and unobserved ($x(s^*)$) as well as the location errors, $\bu_n$. This is in contrast to usual interval estimates from Gaussian process regression without location error, which are exact probability statements conditional on the observed data $\bx_n$. The reason this is an important distinction is because the usual Gaussian process conditional probability intervals yield the proper coverage rate across multiple prediction intervals (when predicting $x$ at a collection of unobserved locations $\bs^*_k$), whereas the confidence intervals corresponding to KALE may not.

\subsection{Advantages over Kriging while Ignoring Location Errors}

Failing to adjust for location errors when Kriging (\citet{Cressie2003} called this ``Kriging ignoring location errors'' or KILE) can lead to poor performance. A data analyst ignoring the location errors will use (see Equation \eqref{GPconditional})
\begin{equation}\label{xkriging}
\xkile(s^*) = \C(s^*, \bs_n)\C(\bs_n, \bs_n)^{-1} \by_n.
\end{equation}
%a procedure we will refer to as $x$-\textit{Kriging} due to the coefficients $\C(s^*, \bs_n)\C(\bs_n, \bs_n)^{-1}$ deriving from the second-moment properties of $x$. This contrasts with the correct Kriging equations \eqref{ykriging} which we shall henceforth refer to as $y$-\textit{Kriging} to emphasize this distinction.
Since $\xkale(s^*)$ is the best linear unbiased estimator for $x(s^*)$ and $\xkile$ is also an unbiased linear estimator, KALE dominates KILE and always yields a reduced MSE. Figure \ref{fig:ignore_mspe_plot} illustrates the disparity in MSE for a simple model; intuitively, the relative cost of ignoring location errors increases as the magnitude of the location errors increases. We also see in panel (B), illustrating Proposition \ref{prop1}, that for small values of $\sigma^2_u$, the MSE for both KALE and KILE decreases in $\sigma^2_u$. 

\begin{knitrout}
\definecolor{shadecolor}{rgb}{0.969, 0.969, 0.969}\color{fgcolor}
\begin{figure}[h!]
\subfloat[Locations at which we observe $y(s)$, as well as the location at which we wish to predict x(s). Sample paths of Gaussian processes with this covariance function are also shown.\label{fig:ignore_mspe_plot1}]{
\includegraphics[width=\maxwidth]{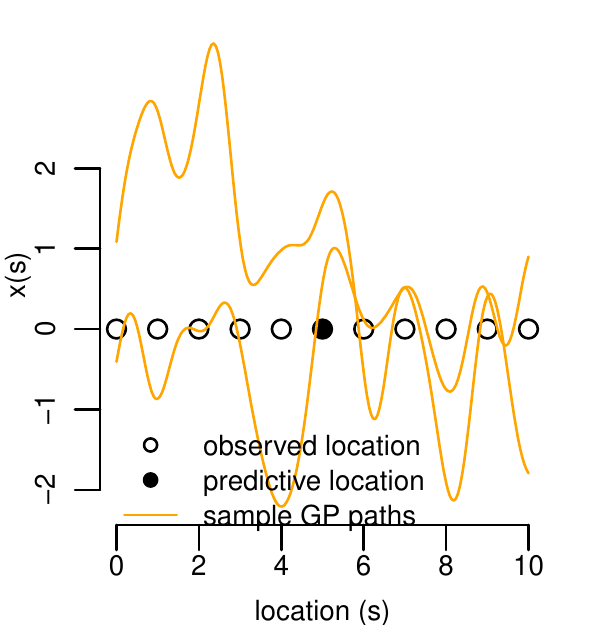} }
\subfloat[For both \emph{KALE} and \emph{KILE}, \emph{MSE} actually decreases as the magnitude of location errors increases when this magnitude is small. Above a certain point, greater location error yields higher \emph{MSE} and greater disparity between \emph{KALE} and \emph{KILE}.\label{fig:ignore_mspe_plot2}]{
\includegraphics[width=\maxwidth]{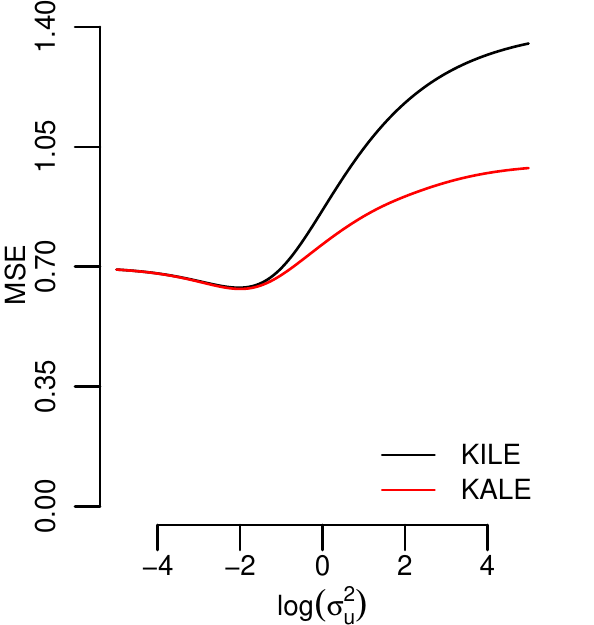} }\caption[Here we assume $x(s)$ is a Gaussian process with mean 0 and covariance function $c(s_1, s_2) = \exp(-(s_1 - s_2)^2)$, with $u_i \stackrel{iid}{\sim} \N(0, \sigma^2_u)$]{Here we assume $x(s)$ is a Gaussian process with mean 0 and covariance function $c(s_1, s_2) = \exp(-(s_1 - s_2)^2)$, with $u_i \stackrel{iid}{\sim} \N(0, \sigma^2_u)$. We compare \emph{MSE} for predicting $x(5)$ based on $y(0), \ldots, y(4)$ using \emph{KALE} and \emph{KILE}.\label{fig:ignore_mspe_plot}}
\end{figure}
\end{knitrout}

%When performing simple Kriging using the usual equations, the presence of location errors may yield higher mean squared predictive error. But this is not always the case. It is actually possible that subjecting the data to location errors---and ignoring them during Kriging---\textit{reduces} mean squared error when predicting $x$ at new locations. Please refer to Figure \ref{fig:ignore_mspe_plot} for a simple example illustrating this.

Besides yielding suboptimal predictions relative to KALE, ignoring location errors also leads to an estimator for $x(s^*)$ that is not self-efficient [\textit{p}. 549, \citet{meng1994multiple}]. Following \citet{meng1994multiple}, an estimator $T$ for parameter $\theta$ is self-efficient if for any $\lambda \in [0,1]$ and subset of the observed data $X_c \subset X$, we have 
$$\E[(\lambda T(X) + (1 - \lambda)T(X_c) - \theta)^2] \geq \E[(T(X) - \theta)^2].$$ 
Thus, roughly speaking, self-efficient estimators are those that cannot be improved by using only a subset of the original data [\citet{meng2014got}].

The following theorem states that the KILE MSE is unbounded as a function of any single spatial location $s_i$ for $i=1, \ldots, n$. This is a stronger result than just the lack of self-efficiency. A consequence of Theorem \ref{theorem1} is that, assuming only simple continuity conditions on the covariance function and location error model, the KILE MSE can always increase when observing more data, regardless of the locations of the existing observations or the locations at which we want to make predictions.
%\begin{theorem} Assume $c$ satisfies regularity conditions A1--A4 (in appendix) and $u \sim g(u)$. Let $\xkile^{n}(s^*)$ be the KILE estimator for $x(s^*)$ given $\bx_n$. Then for any $\bs_n$ and $s^*$, there exists $s_{n+1}$ such that $\E[(x(s^*) - \xkile^{n+1}(s^*))^2] \geq \E[(x(s^*) - \xkile^{n}(s^*))^2]$. \label{theorem1}
%\end{theorem}
\begin{theorem} Suppose that the following conditions hold:
\begin{itemize}
\item $c$ is continuous and bounded in $\Ss^2$,
\item the location error model $g$ satisfies $(u_1^m, u_2^m) \stackrel{D}{\rightarrow} (u_1, u_2)$ for all $s_1, s_2 \in \Ss$ and sequences $(s_1^m, s_2^m)$ such that $\lim_{m \to \infty} (s_1^m, s_2^m) = (s_1, s_2)$, 
\item and $\prob(u_1 \neq u_2) < 1$ for all $s_1, s_2 \in \Ss$.
\end{itemize}
%\item location errors satisfy $\prob(u_1 \neq u_2) < 1$ for all $s_1, s_2 \in \Ss$
Let $\xkile(s^*)$ be the \emph{KILE} estimator for $x(s^*)$ given $\by_n$. Then for any $M > 0$, $n \geq 2$, and $s_2, \ldots, s_n \in \Ss$, there exists $s_1$ such that $\E[(x(s^*) - \xkile(s^*))^2] > M$. \label{theorem1}
\end{theorem}
%\end{theorem}
%\todo{Analogous result for nugget?}
Note that the condition that $c$ is continuous excludes a nugget term from the distribution of $x$. We prove Theorem \ref{theorem1} (in the Appendix) by showing that when observed locations are very close together, the corresponding covariance matrix is nearly singular, and this increases MSE. Without location errors, the usual Kriging estimator does not exhibit this behavior since the difference between values of $x(s)$ for points that are close together also converges in probability to 0. This is not the case for the noise-corrupted process, as $y(s_2) - y(s_1)$ does not converge to 0.

\begin{knitrout}
\definecolor{shadecolor}{rgb}{0.969, 0.969, 0.969}\color{fgcolor}\begin{figure}[H]

\includegraphics[width=\maxwidth]{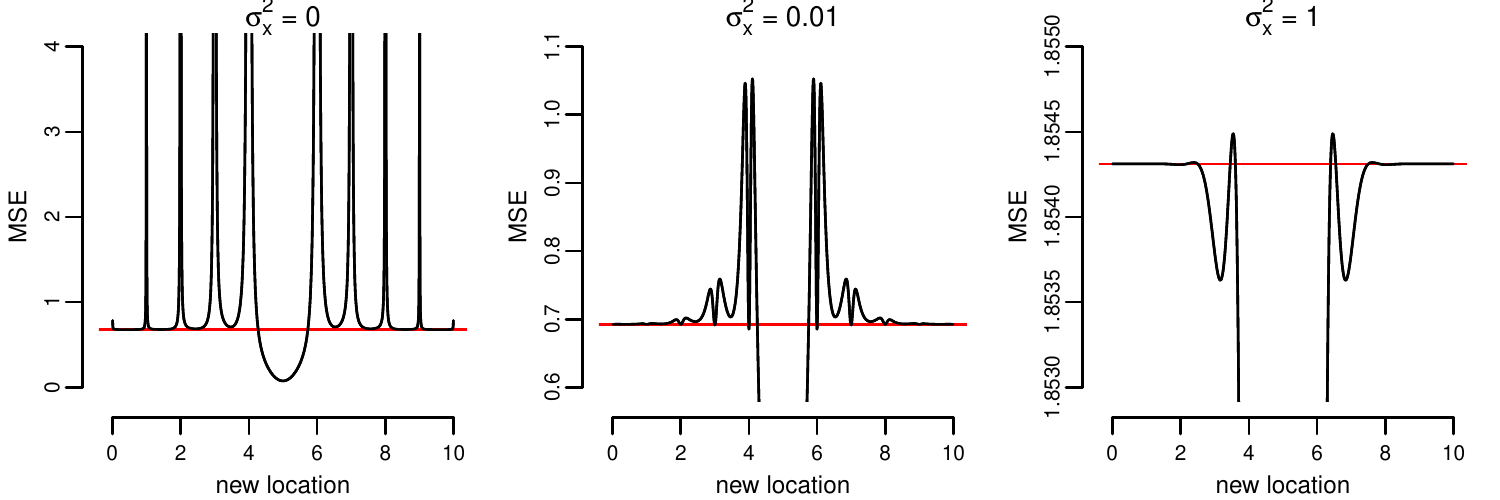} \caption[Here we assume $x(s)$ is a Gaussian process with mean 0 and covariance function $c(s, s^*) = \exp(-(s - s^*)^2) + \sigma^2_x \mathbf{1}_{s = s^*}$]{Here we assume $x(s)$ is a Gaussian process with mean 0 and covariance function $c(s, s^*) = \exp(-(s - s^*)^2) + \sigma^2_x \mathbf{1}_{s = s^*}$. Location errors have the form $u_i \sim \N(0, 0.04)$. We use \emph{KILE} to predict $x(5)$ based on $\by_{\text{obs}} = \{y(0), \ldots, y(4), y(6), \ldots, y(10)\}$, as well as an additional observation $y_(s)$. The MSE in predicting $x(5)$ given $\by_{\text{obs}}$ and $y(s)$ is plotted as a function of $s$, while the red line denotes 
the \emph{MSE} based only on $\by_{\text{obs}}$. Despite the magnitude of the location errors being relatively small, observing another measurement of $y$ at some locations can increase (possibly dramatically) the \emph{MSE}.\label{fig:no_self_efficiency}}
\end{figure}

\end{knitrout}

Simulation results suggest that even when $c$ contains a nugget term $\sigma^2_x$, KILE is still not self-efficient, and additional observations can increase MSE. Figure \ref{fig:no_self_efficiency} illustrates the change in MSE as a function of the location of an additional observation of $y$. Following Theorem \ref{theorem1}, we see the MSE is unbounded when $\sigma^2_x = 0$. But even when $\sigma^2_x = 1$, it is possible for an additional observation to (slightly) increase MSE.

%\todo[inline]{Another subsection on computational issues with $y$-Kriging here?}

\subsection{Parameter Estimation for Kriging}

%identifiability with $\sigma^2_u, micro-scale variation??

In typical applied settings, some or all parameters of the covariance function are unknown and must be estimated by the analyst in order to obtain Kriging equations. For Gaussian process models without a location error component, parameter estimation can be accomplished using likelihood methods. This can be computationally challenging for large data sets, as each likelihood evaluation requires a Cholesky factorization of the covariance matrix (or equivalent operations), which is $\mathcal{O}(n^3)$ except in special cases. An alternative is to choose parameters by maximizing goodness of fit between the empirical variogram and the theoretical (parametric) variogram, though this is less efficient for parametric Gaussian models.

Location errors present challenges for both such procedures as the covariance function for the observed provess $y$ \eqref{cov_y} may not be available in closed form, meaning neither the likelihood function or variogram can be evaluated exactly. While Monte Carlo methods surely offer effective approaches in theory [\citet{fanshawe2011spatial}], they muliply the computational expense of the problem, as each evaluation of the likelihood requires $M$ matrix factorizations, where $M$ is the number of Monte Carlo samples used to approximate the likelihood. \citet{Cressie2003} advocate a pseudo-likelihood procedure [\cite{carroll2006measurement}] that uses a Gaussian likelihood approximation based on the first two moments of $y$,
\begin{equation}\label{pseudo}
\tilde{L}(\theta; \by_n) \propto |\K_{\theta}(\bs_n, \bs_n)|^{-1/2}\exp \left( - \frac{1}{2} \by_n ' \K_{\theta}(\bs_n, \bs_n)^{-1} \by_n \right),
\end{equation}
where we write $\K_{\theta}$ to explicitly mark the dependence of the covariance function $k$ on unknown parameters $\theta$. This pseudo-likelihood requires inverting $\K$ only once per pseudo-likelihood evaluation, even when $\K_{\theta}$ is computed by Monte Carlo. 

We can work out inferential properties of the maximum pseudo-likelihood estimator $\hat{\tilde{\theta}} = \text{argmax}_{\theta} \tilde{L}(\theta; \by_n)$. First, it is straightforward to check that the pseudo-score pertains to an unbiased estimating equation:
\begin{equation} \label{pseudoscore}
\E[\tilde{S}(\theta; \by_n)] = \E[\nabla \log( \tilde{L}(\theta; \by_n))] = \mathbf{0}.
\end{equation}
Moreover, one can show the covariance matrix of the pseudo-score is given by 
\begin{align} \label{varscore}
\tilde{G}(\theta) =& \E[\tilde{S}(\theta; \by_n) \tilde{S}(\theta; \by_n)'] \nonumber \\
\tilde{G}(\theta)_{ij} =& \E \left[\frac{1}{2}\text{Tr}\{\Omega_i \C_{\theta}(\bu_n) \Omega_j \C_{\theta}(\bu_n)\}\right] 
\nonumber \\
& + \frac{1}{4}\Big( \E[ \text{Tr}\{\Omega_i \C_{\theta}(\bu_n)\}\text{Tr}\{\Omega_j \C_{\theta}(\bu_n)\} ] - \text{Tr}\{\Omega_i \K_{\theta}\}\text{Tr}\{\Omega_j \K_{\theta}\} \Big),
\end{align}
using the notational abbreviations $\C_{\theta}(\bu_n) = \C_{\theta}(\bs_n + \bu_n, \bs_n + \bu_n)$, $\K_{\theta} = \K_{\theta}(\bs_n, \bs_n) = \E[\C_{\theta}(\bu_n)]$, and 
%\begin{equation*}
$\Omega_i = \K_{\theta}^{-1} \left(\frac{\partial}{\partial \theta_i} \K_{\theta} \right) \K_{\theta}^{-1}$. 
%\end{equation*}
Lastly, the expected negative Hessian of the log pseudo-likelihood is
\begin{align} \label{hessian}
\tilde{H}(\theta)_{ij} &= \E \left[- \frac{\partial^2}{\partial \theta_i \partial \theta_j} \log( \tilde{L}(\theta; \by_n)) \right] \nonumber \\
 &= \frac{1}{2}\text{Tr}\{\Omega_i \K_{\theta} \Omega_j \K_{\theta}\}.
 \end{align}

If there are no location errors ($\bu_n \equiv \mathbf{0}$), $\tilde{L}$ is an exact likelihood and the second term in the right hand side of Equation \eqref{varscore} vanishes so that $\tilde{G}(\theta) = \tilde{H}(\theta)$, confirming the second Bartlett identity [\citet{ferguson1996course}]. For non-zero location errors, however, we construct the Godambe information matrix as an analog to the Fisher information matrix [\citet{varin2011overview}], 
$$\tilde{I}(\theta) = \tilde{H}(\theta) [\tilde{G}(\theta)]^{-1} \tilde{H}(\theta).$$ Evaluating $\tilde{I}(\theta)$ for different location error models illustrates the information loss in estimating covariance function parameters $\theta$ relative to the error-free case, where $\tilde{I}(\theta) = \tilde{G}(\theta) = \tilde{H}(\theta)$ is equivalent to the Fisher information matrix. 
%\todo[inline]{Provide numerical example?}

General theory of unbiased estimation equations [\citet{heyde1997quasi}] suggests the asymptotic behavior of the pseudo-likelihood procedure satisfies
\begin{equation}
\tilde{I}(\theta)^{1/2}(\hat{\tilde{\theta}} - \theta) \stackrel{D}{\rightarrow} \N(\mathbf{0}, \mathbf{I}).
\label{asym}
\end{equation}
However, Expression \eqref{asym} does not hold in general even in an error-free regime $\bu_n \equiv 0$, as asymptotic results for Gaussian process covariance parameters depend on the spatial sampling scheme used and the specific form of the covariance function [\cite{stein1999interpolation}]. We nevertheless expect \eqref{asym} to hold for suitably well-behaved processes under increasing-domain asymptotics. \cite{guyon1982parameter} gives an applicable result when locations $\bs_n$ are on a lattice. We are not aware of other theoretical results in this context. %{\tcr{Is this ok?}} {\dc{I think so}}

\section{Markov Chain Monte Carlo Methods}
\label{sec:hmc}

%C. Monte Carlo inference
% 1. requires distributional assumption; gaussian makes sense.
% 2. information in location error, mutual information in location error and prediction
% 3. known parameters: perfect conditional coverage, dominates KALE
% 4. unknown parameters: obviously extends with hmc, choices of prior
% 5. computational/convergence issues (reducing parameter space)

Markov Chain Monte Carlo methods offer an alternative to Kriging for prediction in a regime with noisy inputs. They allow us to compute the MSE-optimal prediction
\begin{align}
\hat{x}(s^*) &= \E[x(s^*) | \by_n ] \nonumber \\
 &=\int \left( \C(s^*, \bs_n + \bu_n)[\C(\bs_n + \bu_n, \bs_n + \bu_n)]^{-1}\by_n \right) \pi(\bu_n | \by_n)d\bu_n,
\label{bayesrule}
\end{align}
which will dominate the KALE estimator \eqref{ykriging} in terms of MSE for any model and set of observed and predicted locations. The optimality of $\hat{x}(s^*)$ in \eqref{bayesrule} is due to the fact that the conditional mean $\E[x(s^*) | \by_n]$ obtains the minimum MSE for any estimator of $x(s^*)$ that is a function of $\by_n$. This estimator is not linear, and MCMC methods are necessary for evaluating \eqref{bayesrule} as the density for the conditional distribution $\pi(\bu_n | \by_n)$ will not be available in closed form (no possible ``conjugate`` form for the distribution of $\bu_n$ is known to us). When model parameters, such as in the covariance function $c$ or the distribution of $u$ are unknown, the distribution $\pi(\bu_n | \by_n)$ implicitly averages over the posterior distributions of such parameters.

MCMC methods also allow us to compute prediction intervals $(z_{\text{low}}, z_{\text{high}})$ such that $\prob(z_{\text{low}} < x(s^*) < z_{\text{high}} | \by_n) = 1 - \alpha$. When the covariance function $c$ and location error model $g$ are known, these intervals are exact probability conditional probability statements, providing a stronger coverage guarantee than that achieved with the KALE procedure in Proposition \ref{coverprop}, where coverage is achieved only by averaging over $\by_n$.

%{(\tcr{I don't quite understand this; Bayesian methods don't guarantee coverage!})} {\dc{The MCMC procedure described here is not a Bayesian (inference) method, it's only invoked for sampling/integration purposes}}

\subsection{Distributional Assumptions}

MCMC inference for \eqref{bayesrule} requires the assumption that $\bx_n$ is Gaussian. While this is a common assumption in practice and has been assumed throughout this paper, it is not necessary to derive the KALE equations and their MSE (but it is necessary to produce coverage intervals as in Proposition \ref{coverprop}). Thus, Kriging approaches, including KALE, are attractive when there is information about the joint distribution of $x$ beyond its first two moments.

In this scenario, however, we can still advocate---from a decision-theoretic perspective---a Gaussian assumption when the goal of the analysis is minimum MSE prediction. 
%Our reasoning follows that of \cite{morris1983natural}, who shows the merits of using conjugate priors for hierarchical models where the prior mean and variance are fixed. For hierarchical models, assuming the conjugate prior as the distributional form of the model parameters means that the MSE only depends on prior through its first two moments; this is not true for other choices of prior. When the first two prior moments are fixed, the conjugate prior is a conservative choice of distribution because the resulting Bayes rule estimator has the same MSE under any true prior (Bayes risk) {\tcr{change this!}} {\dc{I don't understand what the problem is}} with the same first two moments. 
%Since in Kriging, we specify the first two moments of $\bx_n$, the above discussion suggests that the Gaussian assumption makes sense as a conservative choice of joint distribution for $\bx_n$. 
Specifically, let $\pi \in \Pi_{\mathbf{0}, \C}$ be a choice of joint distribution for $\bx_n$ with the appropriate first two moments $\mathbf{0}$ and $\C$. 
%Let $\hat{x}_{\pi}(s^*; \bx_n)$ be an estimator for $x(s^*)$ based on $\bx_n$ and the assumption that $\bx_n \sim \pi$. Because conditional means minimize MSE, it makes sense to simply assume $\hat{x}_{\pi}(s^*; \bx_n) = \E_{\pi}[x(s^*) | \bx_n]$
The minimum MSE prediction of $x(s^*)$ assuming $\bx_n \sim \pi$ is the conditional mean $\E_{\pi}[x(s^*) | \bx_n]$. Let $\mathrm{R}_{\pi_0}(\pi)$ be the risk (MSE) of this minimum MSE predictor under the assumption that $\bx_n \sim \pi$ when in fact $\bx_n \sim \pi_0$; that is, 
$$\mathrm{R}_{\pi_0}(\pi) = \E_{\pi_0}[(\E_{\pi}[x(s^*) | \bx_n] - x(s^*))^2].$$
Thus $R_{\pi_0}(\pi)$ represents the risk (MSE) in a misspecified joint distribution for $\bx_n$, where the analyst assumes $\bx_n \sim \pi$, but in fact $\bx_n \sim \pi_0$. We then have the following proposition, based on Theorem 5.5 of \cite{morris1983natural}:
\begin{proposition} Let $\pi_0 \in \Pi_{\mathbf{0}, \C}$ be Gaussian. Then for all $\pi \in \Pi_{\mathbf{0}, \C}$ we have
$$\mathrm{R}_{\pi}(\pi) \leq \mathrm{R}_{\pi}(\pi_0) = \mathrm{R}_{\pi_0}(\pi_0) \leq \mathrm{R}_{\pi_0}(\pi).$$
\label{morris_prop}
\end{proposition}
Unlike traditional decision theory problems, here we are fixing the estimator (Kriging), and considering the costs of different distributional assumptions ($\pi$). Given that the analyst has decided to use Kriging for predicting $x(s^*)$, then the risk in making an incorrect distributional assumption is $\mathrm{R}_{\pi_0}(\pi_0) - \mathrm{R}_{\pi}(\pi_0) = 0$. This reflects the fact that the Kriging MSE depends only on the first two moments of $\pi$. However, there is an ``opportunity cost'' in making any non-Gaussian assumption $\mathrm{R}_{\pi}(\pi_0) - \mathrm{R}_{\pi}(\pi) > 0$ for $\pi \neq \pi_0$, which represents the reduction in MSE under $\pi$ that could be achieved by using a different estimator other than Kriging. 

Obviously, if there is a strong reason to believe a non-Gaussian $\pi$ is true, then analysis should proceed with this assumption, ideally leveraging an estimator that is optimal under these assumptions (instead of Kriging). However, without strong distributional knowledge, the analyst can assume Gaussianity without risking increased MSE or paying an opportunity cost for using an inefficient method. 

%As this also represents a Bayesian inference problem, we distinguish between the known and unknown model parameter paradigms in the discussion to follow.

\subsection{Hybrid Monte Carlo}

Hybrid Monte Carlo [\cite{neal2005hamiltonian}] is well-suited for the problem of sampling $\pi(\bu_n | \by_n) \propto \pi(\by_n | \bu_n) \pi(\bu_n)$ in order to evaluate \eqref{bayesrule}. This is because while $\pi(\bu_n | \by_n)$ is computationally expensive (requiring inversion of the covariance matrix $\Cun = \C_{\theta}(\bs_n + \bu_n, \bs_n + \bu_n)$), the gradient $\nabla \log (\pi(\bu_n | \by_n))$ is a relatively cheap byproduct of this calculation. Often the conditional distribution $\bu_n | \by_n$ is correlated across components, making gradient-based MCMC methods more efficient for generating samples. Other gradient-based MCMC sampling methods, such as the Metropolis-adjusted Langevin algorithm [\citet{roberts2001optimal}] and variants, may also be well-suited to this problem. 

Bayes rule provides $\pi(\theta, \bu_n | \by_n) \propto \pi(\by_n | \theta, \bu_n)\pi(\theta, \bu_n)$, where $\theta$ here represents any unknown parameter(s) of the covariance function $c$. In most situations it will be reasonable to assume $\bu_n$ and $\theta$ are independent a priori---this is trivially true in the case that $\theta$ is assumed known. Assuming this, and recognizing that $\pi(\by_n|\theta, \bu_n)$ is Gaussian, we can write the log posterior and its gradient:
\begin{align*}
\log(\pi(\theta, \bu_n | \by_n)) &= - \frac{1}{2} \log(|\Cun|) - \frac{1}{2} \by_n' \Cun^{-1} \by_n + \text{const.}\\
\ddui \log(\pi(\theta, \bu_n | \by_n)) &= \\
& \hspace{-3cm} \frac{1}{2}\text{Tr}\left(\Cun^{-1} \left[ \ddui \Cun \right] \left(\Cun^{-1} \by_n \by_n' - \mathbf{I}_n \right) \right) + \ddui \log(\pi(\bu_n)) \\
\ddth \log(\pi(\theta, \bu_n | \by_n)) &= \\
& \hspace{-3cm} \frac{1}{2}\text{Tr}\left(\Cun^{-1} \left[ \ddth \Cun \right] \left(\Cun^{-1} \by_n \by_n' - \mathbf{I}_n \right) \right) + \ddth \log(\pi(\theta)).
\end{align*}

The computational cost of both the likelihood and gradient are dominated by solving $\Cun$ (\textit{e.g.}, Cholesky factorization), which is $\mathcal{O}(n^3)$. Every likelihood evaluation computes this term, which can then be re-used in the gradient equations. Thus, the computational cost of computing both the likelihood and gradient remains $\mathcal{O}(n^3)$.

\subsection{Multimodality}

The posterior distribution $\pi(\theta, \bu_n | \by_n)$ is often multimodal, more so if the distrubution $\pi(\bu_n)$ is diffuse. This is because if there is a local mode at $(\hat{\theta}, \hat{\bu}_n)$, there may be a local mode at any $(\theta, \bu_n)$ such that $\Cun = C_{\hat{\theta}}(\hat{\bu}_n)$, as the likelihood is constant for such $(\theta, \bu_n)$. In particular, for isotropic covariance models, the likelihood is constant for additive shifts in $\bu_n$ or rotations of $\bs_n + \bu_n$, as these operations preserve pairwise distances. Additionally, multimodality can be induced by the many-to-one mapping of the set of true locations $\{s_i + u_i, i=1, \ldots, n\}$ to the set of observed locations $\{s_i, i=1, \ldots, n\}$. For instance, with $n=2$ and an isotropic covariance function, for any choice of $u_1, u_2$ we get the same likelihood with $\tilde{u}_1 = s_2 + u_2 - s_1$ and $\tilde{u}_2 = s_1 + u_1 - s_2$. Moreover, for fixed $\bu_n$, for many common covariance functions it is possible for the posterior of $\theta$ to be multimodal [\citet{warnes1987problems}].

HMC (and other gradient MCMC methods) can efficiently sample from multiple modes, however this becomes difficult when the modes are isolated by regions of extremely low likelihood [\citet{neal2011mcmc}]. Isolated modes can occur in the location-error GP regime. For example, assume one-dimensional locations ($p=1$) and an isotroptic covariance model with known parameters $\theta$ and nugget $\sigma^2_x$. Marginally, as $ \| s_1 + u_1 - (s_2 + u_2)\| \rightarrow 0$, $y_1 - y_2 \stackrel{D}{\rightarrow} \N(0, 2\sigma^2_x)$; that is, the scaled difference $|y_1 - y_2|/(\sigma_x)$ must be reasonably small. When this is not the case (e.g. $\sigma^2_x = 0$), then the log-likelihood asymptotes at $s_1 + u_1 = s_2 + u_2$ almost surely. Thus, the Markov chain can only sample $\bu_n$ such that the ordering of $\{s_i + u_i, i=1, \ldots, n\}$ is preserved. Note that when $p > 1$, while the log-likelihood may still asymptote at $s_1 + u_1 = s_2 + u_2$, this no longer constrains the space of $\bu_n$ (except on sets with measure 0).

\begin{knitrout}
\definecolor{shadecolor}{rgb}{0.969, 0.969, 0.969}\color{fgcolor}\begin{figure}[H]
\subfloat[$\sigma^2_u=2$, $\sigma^2_x = 0.0001$ \label{fig:multi_modes_plot1}]{
\includegraphics[width=\maxwidth]{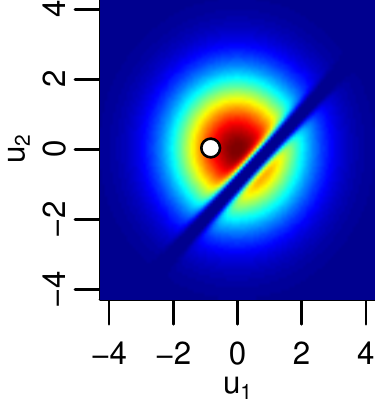} }
\subfloat[$\sigma^2_u=2$, $\sigma^2_x = 1$\label{fig:multi_modes_plot2}]{
\includegraphics[width=\maxwidth]{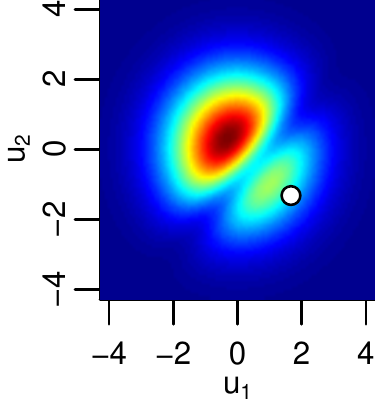} }
\subfloat[$\sigma^2_u=0.1$, $\sigma^2_x = 0.0001$\label{fig:multi_modes_plot3}]{
\includegraphics[width=\maxwidth]{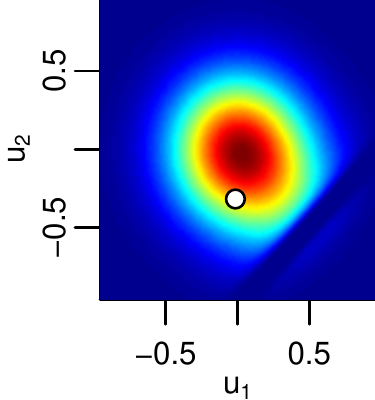} }\caption[Density of $(u_1, u_2)$ using the covariance function ${c(s_1, s_2) = \exp(-(s_1 - s_2)^2)}$ and nugget $\sigma^2_x$]{Density of $(u_1, u_2)$ using the covariance function ${c(s_1, s_2) = \exp(-(s_1 - s_2)^2)} + \sigma^2_x\mathbf{1}_{s1 = s2}$. We simulate data $(y_1, y_2)$ using $s_1 = 0$, $s_1 = 1$, and $u_i \sim \N(0, \sigma^2_u)$, and different values of $\sigma_x^2$ and $\sigma_u^2$.\label{fig:multi_modes_plot}}
\end{figure}

\end{knitrout}

Figure \ref{fig:multi_modes_plot} demonstrates the modal behavior for this simple example with $p=1$ and $n=2$. When location errors are large in magnitude and the nugget tern $\sigma^2_x$ is small, the modes of $(u_1, u_2)$ are separated by a contour of near 0 density (panel A). A higher nugget $\sigma^2_x$ increases the density between the modes, making it easier for the same MCMC chain to travel between them (panel B). Decreasing the magnitude of the (Gaussian) location errors, $\sigma^2_u$, puts more mass on a single mode, as the unimodal distribution $\pi(\bu_n)$ has a greater influence on $\pi(\bu_n | \by_n)$ (panel C).

Thus, as with any MCMC application, for the location-error GP problem it is advisible to run separate chains in parallel, with different, diffuse starting points, and monitor mixing diagnostics [\citet{gelman2011inference}]. Multiple chains failing to mix is likely a symptom of multiple isolated modes, in which case we should modify the HMC algorithm to include tempering [\citet{salazar1997simulated}] or non-local proposals that allow for mode switching [\citet{qin2001multipoint, lan2013wormhole}]. Another strategy to overcome multiple isolated modes is importance sampling: as Figure \ref{fig:multi_modes_plot} shows, increasing the nugget variance $\sigma_x^2$ increases the density between modes. If we generate samples according to $\tilde{\pi}(\theta, \bu_n | \by_n) \propto \tilde{\pi}(\by_n | \theta, \bu_n) \pi(\theta) \pi(\bu_n)$ where $\tilde{\pi}(\by_n | \theta, \bu_n)$ is the density corresponding to $\N(\bo, \Cun + \kappa \mathbf{I}_n)$ for some fixed $\kappa$, then it is straightforward to compute importance weights $\pi(\theta, \bu_n | \by_n) / \tilde{\pi}(\theta, \bu_n | \by_n)$. This is because $\Cun^{-1}$ is easy to compute from $(\Cun + \kappa\mathbf{I}_n)^{-1}$ (and vice versa) using the Woodbury  formula. Either standard importance sampling, or Hamiltonian importance sampling [\citet{neal2005hamiltonian}], could be used to generate parameter estimates, point/interval predictions, and any other posterior estimates of interest.

%Another strategy to improve the efficiency of sampling among different modes, is to use importance sampling. To overcome the lack of mixing in our 1 dimensional example above, we could generate samples according to a density $\tilde{pi}$, based on using the same form of $\pi$ but with a higher magnitude nugget term $\tilde{\sigma}^2_x >> \sigma^2_x$. Using, $\tilde{\pi}$ to generate samples $\{\bu_n\}_1, \ldots, \{\bu_n\}_M$, it is easy to compute importance weights $\pi(\bu_n)/\tilde{\pi}(\bu_n)$ that can be used in point/interval estimates.

% HMC crossing wells
% multimodality as a function of Sigma

% provide gradient?
% identifiability, choice of prior
% computational/convergence issues

%\subsection{Known model parameters}

%When all model parameters are known, the observed (location-error-corrupted) data $\by_n$ provides information for the unobserved location errors $\bu_n$. Good knowledge of $\bu_n$ not only allows for proper uncertainty quantification for predicting new observations $x(s^*)$; it can also yield predictions with lower MSE as if $x(s_i + u_i)$ is more correlated with $x(s^*)$ than is $x(s_i)$, we can take advantage of this to obtain more precise predictions if we know $u_i$. 

\section{Simulation study}
\label{sec:simul}

We compare Kriging (both KALE and KILE) and HMC methods for point/interval forecasts for Gaussian process regression in a simulation study. For various combinations of parameter values for the covariance function $c(s_1, s_2)$ and location error model $g(u)$ we simulate observations $\by_n$ where $y_i = x(s_i + u_i)$ and make predictions for values of $x$ at unobserved locations: $\bx^*_k = (x(s^*_1) \: \: \ldots \: \: x(s^*_k))'$. 

%The additive error terms $\epsilon_i$ induce a nugget term in the covariance matrix for the observed data; however, it is important to note here that by introducing the unobserved error variables $\epsilon_i$, we leave $x(s)$ to represent a noiseless, smooth surface. 

We simulate data using the squared exponential covariance function $c(s_1, s_2) = \tau^2 \exp(-\beta \|s_1 - s_2\|^2) + \sigma^2_x \mathbf{1}_{s_1 = s_2}$ and an i.i.d. Gaussian location error model $u_i \stackrel{iid}{\sim} \N(0, \sigma^2_u\mathbf{I}_p)$. The squared exponential covariance function and Gaussian location error model combine to form a convenient regime, as we can evaluate $k$ in closed form \eqref{cov_y_sqexp}. Without loss of generality, we can use $\tau^2 = 1$ for all simulations as it is simply a scale parameter. We consider a $p=2$ dimensional location space, $s_i \in \R^2$. On a $8 \times 8$ grid, we randomly select 54 locations at which we observe $y$, and target the remaining 10 locations for interpolating $x$. Figure \ref{fig:sim_plot} illustrates a range of data samples for processes used in our simulations on this space, while Table \ref{param_table} provides a full summary of all the parameter value combinations we consider. Data from each parameter combination is simulated 100 times.

\begin{table}
  \begin{center}
  \begin{tabular}{lrr}
		\toprule
		Parameter & Values simulated & Prior support\\
		\midrule
		$\tau^2$ & 1 & $(0, 10)$\\
    $\beta$ & 0.001, 0.01, 0.1, 0.5, 1, 2 & $(0.0005, 3)$\\
    $\sigma^2_x$ & 0.0001, 0.01, 0.1, 0.5, 1 & $(0, 10)$\\
    $\sigma^2_u$ & 0.0001, 0.01, 0.1, 0.5, 1 & $(0, 10)$\\
    \bottomrule
	\end{tabular}
	\end{center}
	\caption[ Parameter values used in simulation study.]{Parameter values used in simulation study. The range $(0.0005, 3)$ for $\beta$ guarantees that at least one pair of points among our observed data has a correlation in the range $(0.05, 0.95)$. This eliminates modes corresponding to white noise processes from the likelihood surface.}
	\label{param_table}
\end{table}

\begin{knitrout}
\definecolor{shadecolor}{rgb}{0.969, 0.969, 0.969}\color{fgcolor}\begin{figure}[H]

\includegraphics[width=\maxwidth]{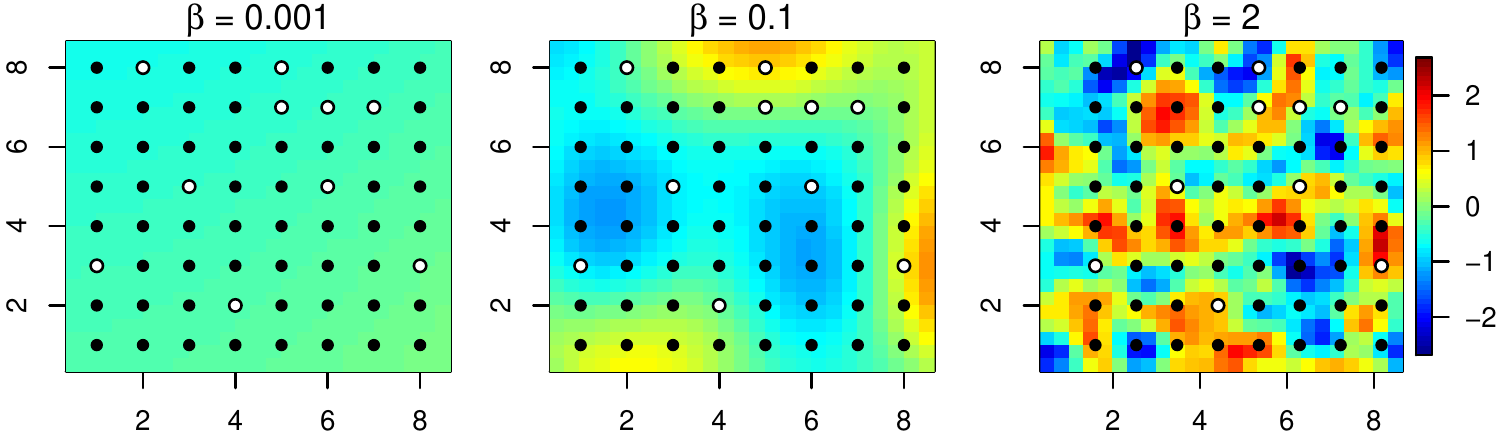} \caption[Samples of $x(s)$ for different values of the length-scale parameter $\beta$ with the squared exponential covariance function, $c(s_1, s_2) = \exp(-\beta \|s_1 - s_2\|^2) + \sigma^2_x \mathbf{1}_{s_1 = s_2}$]{Samples of $x(s)$ for different values of the length-scale parameter $\beta$ with the squared exponential covariance function, $c(s_1, s_2) = \exp(-\beta\|s_1 - s_2\|^2) + \sigma^2_x \mathbf{1}_{s_1 = s_2}$. Black points are where we have observed $y(s)$ and white points are where we wish to predict $x(s)$. Observed/predicted locations were randomly sampled from an $8 \times 8$ grid.\label{fig:sim_plot}}
\end{figure}

\end{knitrout}

We evaluate the three prediction methods---KALE, KILE, and HMC---using both adjusted root mean squared error (RMSE) and the coverage probability of a $95\%$ interval. ``Adjusted'' RMSE is based on the MSE with $\sigma^2_x$ subtracted out, as this term appears in the MSE for any prediction method. For every parameter combination of interest used, these statistics are calculated first by averaging over each of the $k=10$ prediction targets in each simulated draw of new data, and then over the $J=100$ independent data draws. 

Both evaluation statistics can be evaluated more precisely during simulation by utilizing a simple Rao-Blackwellization. For iteration $j$, instead of drawing $\bx^*_k$ in addition to $\by_n$ and calculating $\text{rmse}_j = \|\bx^*_k - \hat{\bx}^*_k\|/k$, we simply condition on the simulated location errors $\bu_n$ to get $\text{rmse}_j = \E[\|\bx^*_k - \hat{\bx}^*_k\|/k \mid \by_n, \bu_n]$. Similarly, to calculate coverage of an interval $(L_{s^*}(\by_n), U_{s^*}(\by_n))$ for $x(s^*)$, for iteration $j$ we use 
$$\text{cov}_j = \frac{1}{k} \sum_{i=1}^k \E[\mathbf{1}[x(s_i^*) \in (L_{s_i^*}(\by_n), U_{s_i^*}(\by_n))] \mid \by_n, \bu_n].$$

HMC is implemented using the software \texttt{RStan} [\citet{rstan-software:2014}], which implements the ``no-U-turn'' HMC sampler [\citet{homan2014no}]. 10000 samples were drawn during each simulation iteration, which (for most parameter values) takes a few minutes on a single 2.50Ghz processor. 

\subsection{Known covariance parameters}

\begin{figure}[h]
\centering
   \subfloat[ \emph{RMSE} ratio of \emph{KALE} to \emph{KILE}.]{\includegraphics{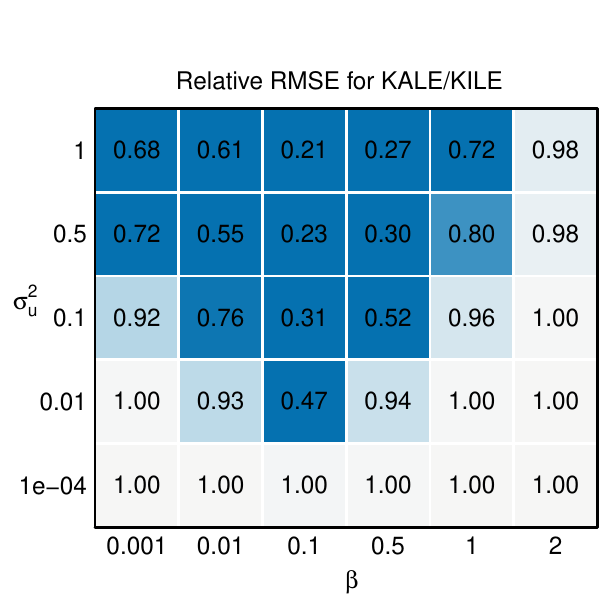}}
  \subfloat[ \emph{RMSE} ratio of \emph{HMC} to \emph{KALE}.]{\includegraphics{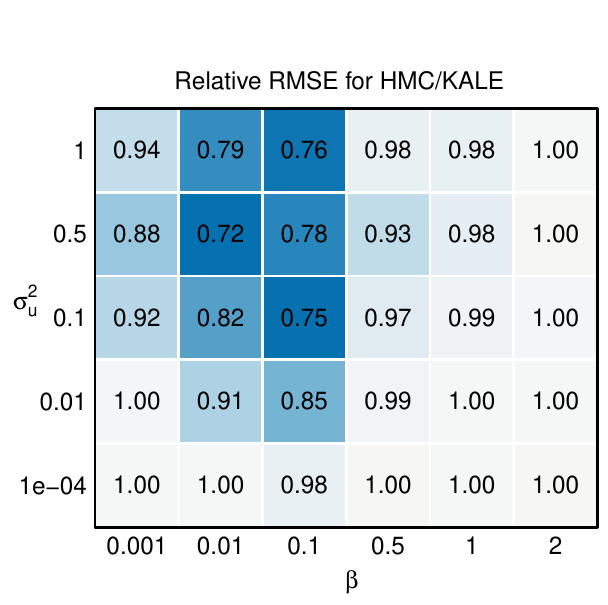}}
\caption[ \emph{RMSE} for \emph{KALE}, \emph{KILE}, \emph{HMC}, parameters known and $\sigma^2_x = 0.0001$]{Relative \emph{RMSE} of \emph{KALE} and \emph{KILE} (A) and \emph{HMC} and \emph{KALE} (B) for each combination of parameters ($\beta$, $\sigma_u^2$) indicated, and $\sigma_x^2 = 0.0001$. Blue shading represents a relative decrease in \emph{RMSE} while red shading represents a relative increase in \emph{RMSE}.} \label{fig:mspe_known_0-0001}
\end{figure}

We first simulate point and interval prediction for KALE, KILE, and HMC using the same parameter values that generated the data. By doing so, we leave aside the issue of parameter inference and simply compare the extent to which different methods leverage the information in the location-error corrupted data $\by_n$ to infer $x(s^*)$. Figure \ref{fig:mspe_known_0-0001} compares RMSE for the three methods when there is a very small nugget, $\sigma_x^2 = 0.0001$. 

We can see that there is little difference among the three methods when $\sigma_u^2$ is sufficiently small ($0.0001$), or when $\beta$ is sufficiently large ($2$). 
This makes sense, as in the former case, with small location errors the potential improvement over KILE (which is exact for $\sigma^2_u = 0$) is negligible, and in the latter case, observations are too weakly correlated for nearby points to be informative. Larger values of $\sigma_u^2$ give KALE a significant reduction in RMSE versus KILE, with the reduction as large as $79\%$ for the case of large magnitude location errors ($\sigma^2_u = 1$) and a moderately smooth signal ($\beta = 0.1$). 

The idea of a moderately smooth signal requires further elaboration: for a given $\sigma^2_u$, when $x$ is very smooth ($\beta$ very small), the process is roughly constant within small neighborhoods, meaning $y(s) \approx x(s)$ and location errors are less of a concern for accurate inference and prediction. On the other hand, when $\beta$ is very large and the process is highly variable in small regions of the input space, location errors are less of a concern because there is very little information in the data to begin with. Location errors are most influential when the process $x$ has more moderate variation across neighborhoods corresponding to the plausible range of the location errors.
%\todo[inline]{Figure illustrating this?}

HMC offers further reductions in RMSE over KALE in roughly the same regions of the parameter space in which KALE improves over KILE, although the additional improvement is less dramatic. The maximum RMSE reduction we observe is about $28\%$, once again for a moderately smooth signal with larger magnitude location errors.

\begin{figure}[h]
\centering
   \subfloat[ \emph{RMSE} ratio of \emph{KALE} to \emph{KILE}.]{\includegraphics{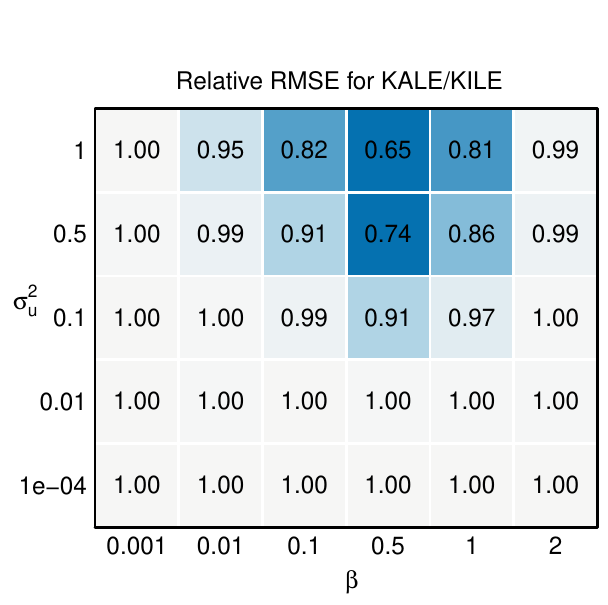}}
  \subfloat[ \emph{RMSE} ratio of \emph{HMC} to \emph{KALE}.]{\includegraphics{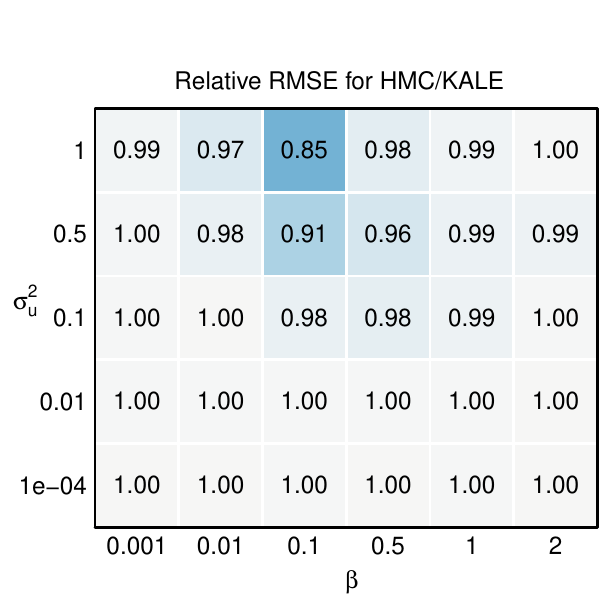}}
\caption[ \emph{RMSE} for \emph{KALE}, \emph{KILE}, \emph{HMC}, parameters known and $\sigma^2_x = 0.1$]{Relative \emph{RMSE} of \emph{KALE} to \emph{KILE} (A) and \emph{HMC} and \emph{KALE} (B) for each combination of parameters ($\beta$, $\sigma_u^2$) indicated, and $\sigma_x^2 = 0.1$. Blue shading represents a relative decrease in \emph{RMSE} while red shading represents a relative increase in \emph{RMSE}.} \label{fig:mspe_known_0-1}
\end{figure}
%\todo[inline]{Better to have one larger figure with all simulation plots?}
When the nugget variance $\sigma^2_x$ is increased (Figure \ref{fig:mspe_known_0-1} shows results for $\sigma^2_x = 0.1$), differences in RMSE among the three methods become smaller (the differences are wiped out entirely at $\sigma^2_x = 1$, which is not pictured). This is not due to a shared $\sigma^2_x$ term in the RMSE value for all methods, as this is subtracted out. Rather, the similarity of all three methods reflects the fact that a larger nugget leaves less information in the data that can be effectively used for prediction. However, the differences that we do observe (both comparing KALE to KILE and HMC to KALE) occur primarily when the magnitude of location errors $\sigma^2_u$ is large.

\begin{figure}[h!]
\centering
   \subfloat[ $\sigma^2_x = 0.0001$.]{\includegraphics{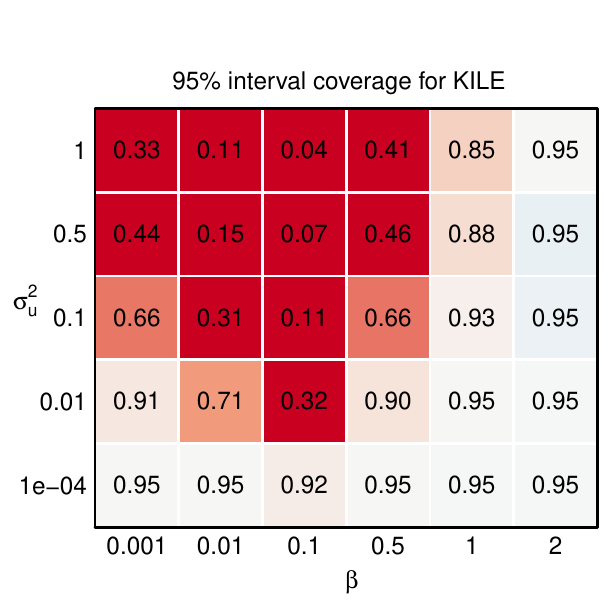}}
  \subfloat[ $\sigma^2_x = 1$]{\includegraphics{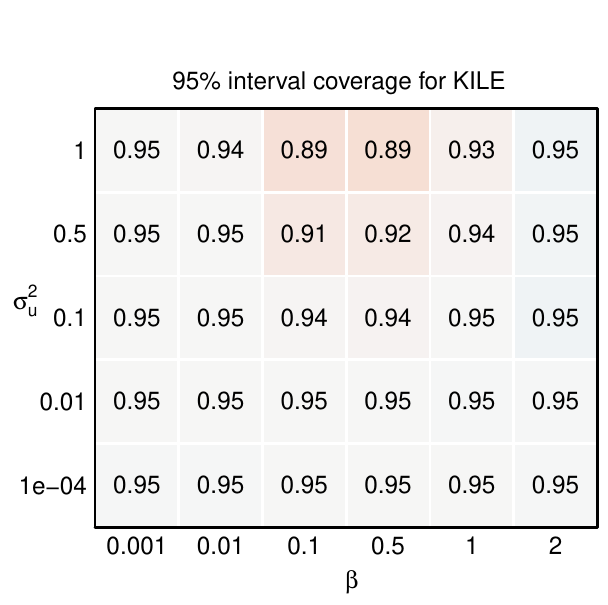}}
\caption[ Interval coverage for \emph{KILE}, parameters known]{$95\%$ interval coverage for \emph{KILE} for $\sigma^2_x = 0.0001$ (A) and $\sigma^2_x = 1$ (B). With moderately smooth signals and large location errors, we see severe undercoverage that does not disappear even for $\sigma^2_x = 1$.} \label{fig:cov_known_none}
\end{figure}

In the case where all parameters are fixed and known, both KALE and HMC produce intervals with exact coverage (subject to Monte Carlo or numerical approximation errors) in all simulations. KILE, however, can severely undercover in the presence of location errors. Figure \ref{fig:cov_known_none} shows coverage as low as $4\%$ when the magnitude of the location errors is high ($\sigma^2_u = 1$), $\beta = 0.1$, and the nugget variation is minimal ($\sigma^2_x = 0.0001$). Undercoverage still persists in this region of the parameter space for $\sigma^2_x = 1$, the largest nugget variance used in our simulations.

\subsection{Unknown covariance parameters}

In typical applied settings, the analyst will not know model parameters such as those of the covariance function ($\tau^2, \beta$), the nugget variance $\sigma^2_x$, or even the variance of the location errors $\sigma^2_u$. Due to identifiability issues with our choice of covariance function in this simulation \eqref{cov_y_sqexp}, we assume $\sigma^2_u$ is known but estimate all other parameters before  making predictions at unobserved locations.

For KILE and KALE, parameter estimation is accomplished through maximum (pseudo-) likelihood, as in \eqref{pseudo}. Parameter estimates are then plugged into Kriging equations \eqref{ykriging}--\eqref{xkriging} to obtain corresponding point and interval estimates. Because $c$ and $k$ are both squared exponential \eqref{cov_y_sqexp}, the pseudolikelihood estimation procedure estimates the same covariance function for $y$, however the estimated parameters (and therefore Kriging equations, based on $k^*$) will differ. The plug-in approach ignores uncertainty in parameter estimates, so plug-in MSE estimates will be too optimistic. Various techniques exist for adjusting MSE from estimated parameters [\citet{smith2004asymptotic, zhu2006spatial}], though there is no need to incorporate such techniques into our analysis since exact (up to Monte Carlo error) MSEs are provided by simulation.

For HMC, we supply unknown parameters with prior distributions and sample parameters and predictions jointly from the posterior distribution $\pi(\theta, \bx^*_k | \by_n)$. The priors we use are flat over a reasonable range (see Table \ref{param_table}), which guarantees both a proper posterior and a posterior mode that agrees with the maximum likelihood estimate of $\theta$. This second condition supports fair comparisons between predictions derived from HMC parameter estimates versus those based on the maximum (psueolikelihood) parameter values. 

\begin{figure}[h!]
\centering
   \subfloat[ \emph{RMSE} ratio of \emph{KALE} to \emph{KILE}.]{\includegraphics{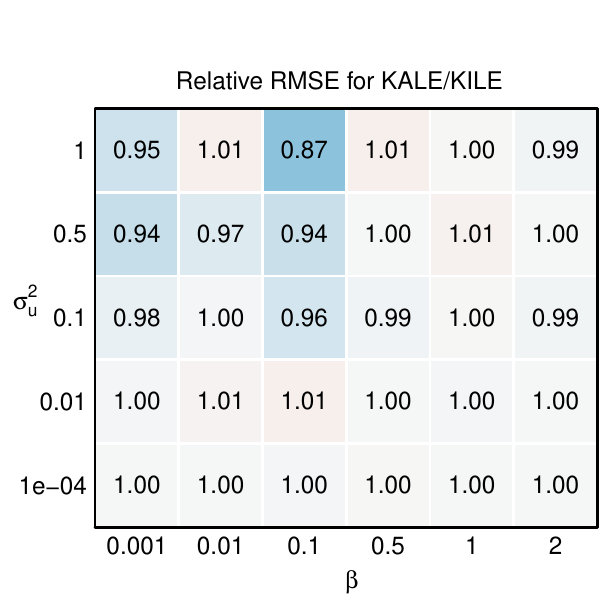}}
   \subfloat[ \emph{RMSE} ratio of \emph{HMC} to \emph{KALE}.]{\includegraphics{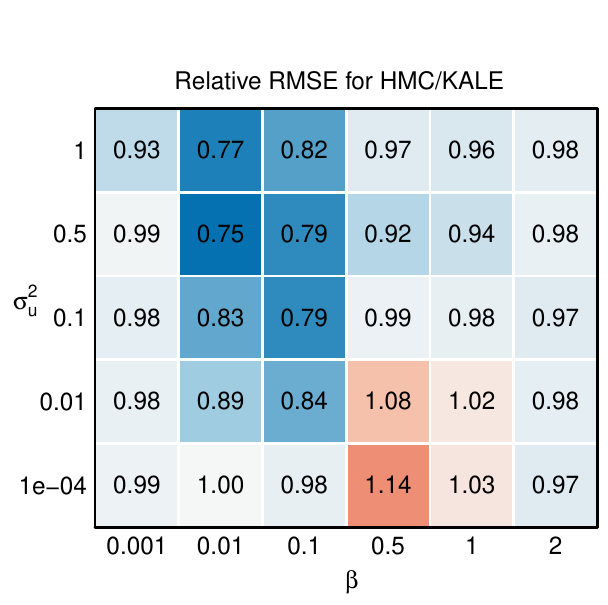}}
\caption[ \emph{RMSE} for \emph{KALE}, \emph{KILE}, \emph{HMC}, parameters unknown and $\sigma^2_x = 0.0001$]{Relative \emph{RMSE} of \emph{KALE} and \emph{KILE} (A) and \emph{HMC} and \emph{KALE} (B) for each combination of parameters ($\beta$, $\sigma_u^2$) indicated, and $\sigma_x^2 = 0.0001$. Parameters are assumed unknown and first estimated to obtain point predictions.}
\label{fig:mspe_unknown_1}
\end{figure}

Figure \ref{fig:mspe_unknown_1} provides the relative RMSE of KALE \textit{vs.} KILE, and HMC \textit{vs.} KALE, for predictions when parameters must first be estimated (using $\sigma_x^2 = 0.0001$). We notice that there does not appear to be a great advantage in KALE over KILE when parameters are first estimated. This is because, as mentioned earlier, the marginal process $y$ still has a squared exponential covariance function \ref{cov_y_sqexp}, so Kriging equations for KALE and KILE will be very similar. On the other hand, we notice a modest improvement when using HMC over Kriging, except in a small region of the parameter space ($\sigma^2_u \leq .01$ and $\beta \in [0.5, 1]$).
\begin{figure}[h!]
\centering
   \subfloat[ \emph{RMSE} ratio of \emph{KALE} to \emph{KILE}.]{\includegraphics{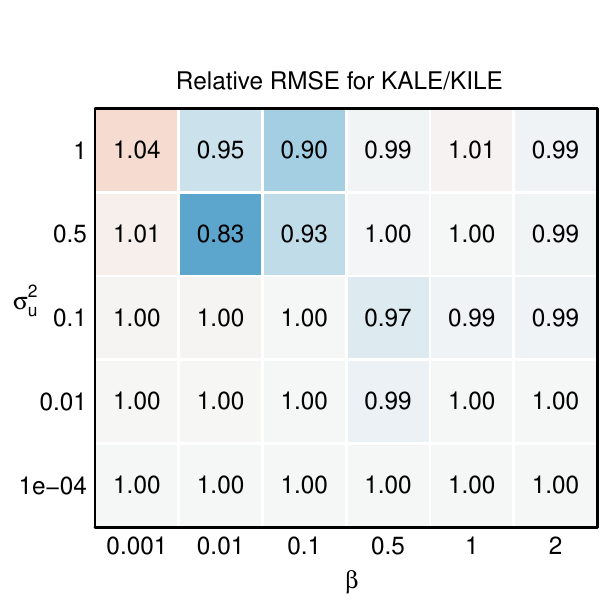}}
   \subfloat[ \emph{RMSE} ratio of \emph{HMC} to \emph{KALE}.]{\includegraphics{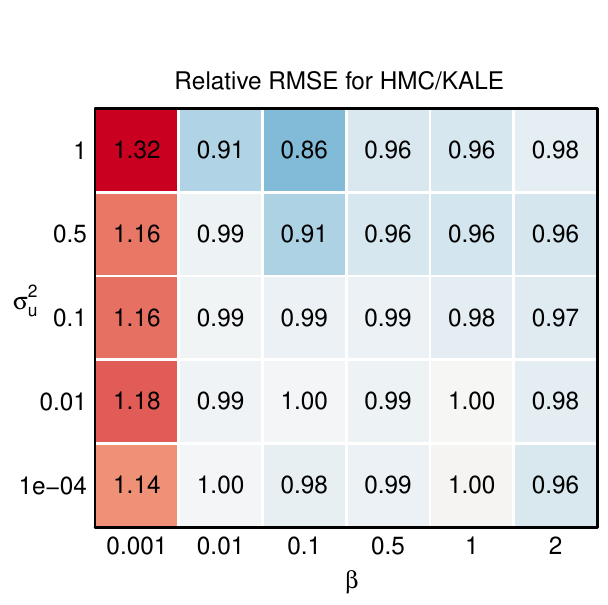}}
\caption[ \emph{RMSE} for \emph{KALE}, \emph{KILE}, \emph{HMC}, parameters unknown and $\sigma^2_x = 0.1$]{Relative \emph{RMSE} of \emph{KALE} and \emph{KILE} (A) and \emph{HMC} and \emph{KALE} (B) for each combination of parameters ($\beta$, $\sigma_u^2$) indicated, and $\sigma_x^2 = 0.1$. Parameters are assumed unknown and first estimated to obtain point predictions.}
\label{fig:mspe_unknown_2}
\end{figure}

When the nugget variance is increased to $\sigma^2_x = 0.1$, we see the results in Figure \ref{fig:mspe_unknown_2}. We still see relatively similar performances from KALE and KILE. HMC offers a small improvement over KALE when $\beta \geq 0.01$, though for $\beta = 0.001$ we actually see significantly higher MSEs with HMC. At $\beta = 0.001$ the process is extremely smooth, as the most distant pairs of observations still have a correlation of $0.88$. We are thus more concerned with overestimating $\beta$ than underestimating it; as the former shrinks predictions towards 0 while the latter shrinks towards (approximately) the mean of all observations. As we use a flat prior for $\beta$, where almost all mass is located $\beta > .001$, the posterior tends to overestimate $\beta$, leading to draws with relatively high MSE.

\begin{figure}[h!]
\centering
   \subfloat[ $\sigma^2_x = 0.0001$]{\includegraphics{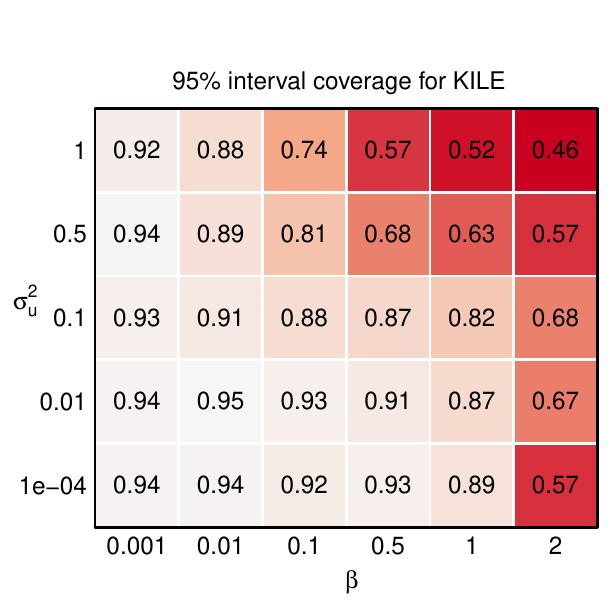}}
  \subfloat[ $\sigma^2_x = 1$ ]{\includegraphics{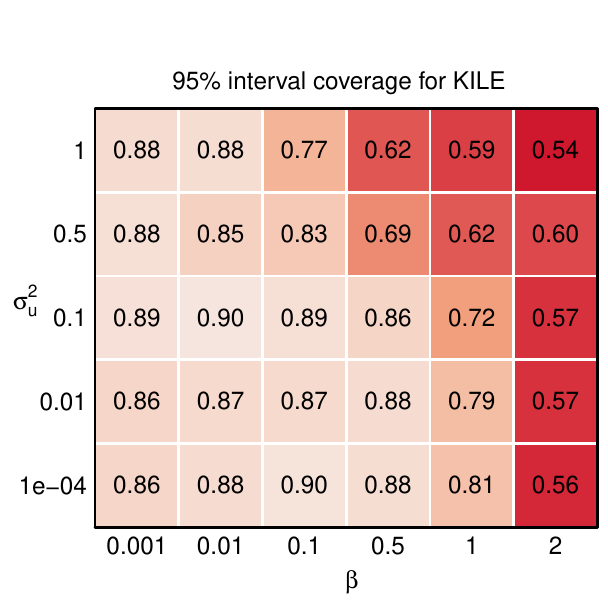}}
\caption[ Interval coverage for KILE, parameters unknown]{$95\%$ interval coverage for \emph{KILE} for $\sigma^2_x = 0.0001$ (A) and $\sigma^2_x = 1$ (B).} 
\label{fig:cov_false_none}
\end{figure}

\begin{figure}[h!]
\centering
   \subfloat[ $\sigma^2_x = 0.0001$]{\includegraphics{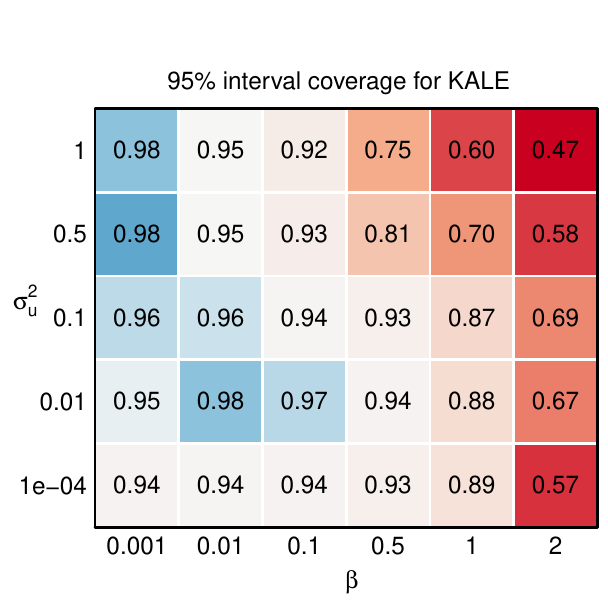}}
  \subfloat[ $\sigma^2_x = 1$ ]{\includegraphics{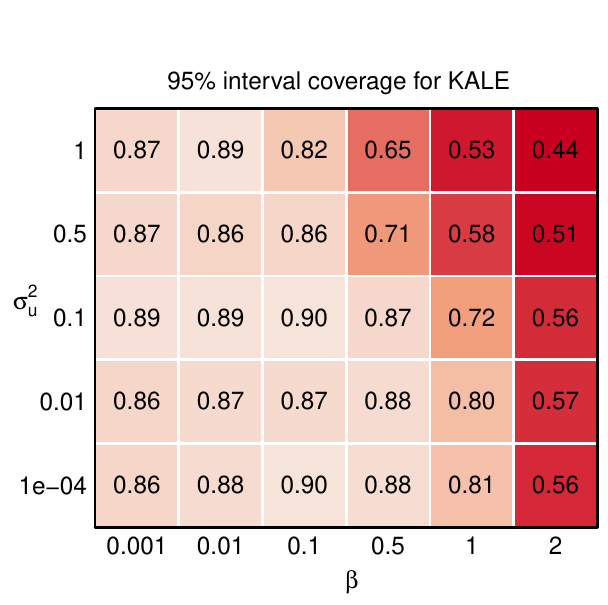}}
\caption[ Interval coverage for KALE, parameters unknown]{$95\%$ interval coverage for KALE for $\sigma^2_x = 0.0001$ (A) and $\sigma^2_x = 1$ (B).} 
\label{fig:cov_false_krig}
\end{figure}

\begin{figure}[h!]
\centering
   \subfloat[ $\sigma^2_x = 0.0001$]{\includegraphics{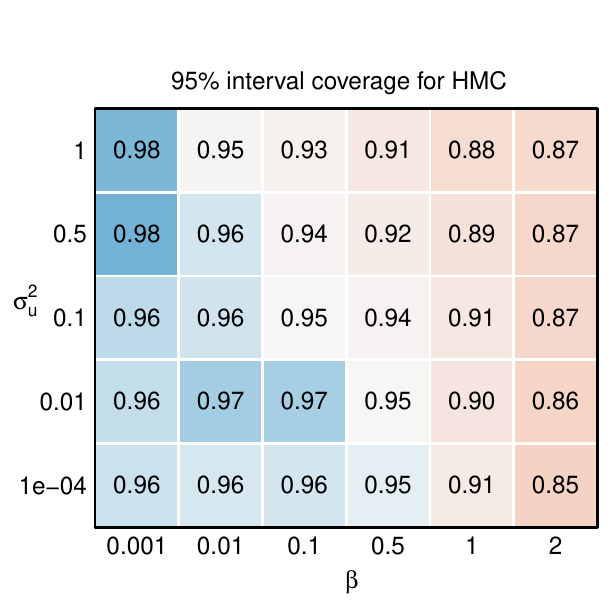}}
  \subfloat[ $\sigma^2_x = 1$ ]{\includegraphics{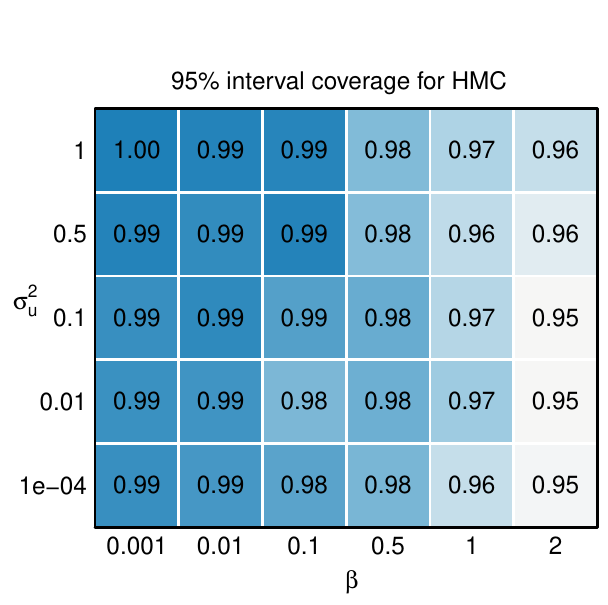}}
\caption[ Interval coverage for HMC, parameters unknown]{$95\%$ interval coverage for HMC for $\sigma^2_x = 0.0001$ (A) and $\sigma^2_x = 1$ (B).} 
\label{fig:cov_false_hmc}
\end{figure}

Neither Kriging or HMC guarantees prediction intervals with the correct coverage in the regime where parameters must first be estimated (though HMC would give proper ``Bayes coverage'' when simulating $\theta$ according to the prior used). We nevertheless present coverage results in Figures \ref{fig:cov_false_none}--\ref{fig:cov_false_hmc}. While we do not expect any method used to provide exact coverage, Kriging (both KALE and KILE) suffer from significant undercoverage for some regions of the parameter space, while HMC is consistent in offering at least $85\%$ coverage throughout our simulations. In a regime without location errors, \citet{zimmerman1992mean} advocate Bayesian procedures under non-informative priors over frequentist procedures in order to obtain interval estimates with good coverage; our simulation results, albeit in the context of location errors, agree with this finding.
%For $x$-Kriging, we estimate parameters using a Gaussian likelihood and squared exponential covariance function, as would be the natural procedure if the data did not have location errors. For $y$-Kriging, we use the pseudo-likelihood \eqref{pseudo} based on the covariance function for $y$ \eqref{y_sq_exp}, which adjusts for location errors. Because of the non-identifiability of ($\tau^2, \beta, \sigma^2_u$), we assume $\sigma^2_u$ is known and set it equal to its true value in these likelihood equations. It is reasonable that such a parameter might be known to applied researchers through instrumental calibration. For convenience, we also assume $\sigma^2_x$ is known, as makes numerical optimization of the likelihood and pseudo-likelihood better behaved and easier to automate during simulation. For both Kriging methods, estimated parameters are plugged into the Kriging equations to make predictions; this will understate predictive uncertainty, yet is common practice in geostatstics as propagating estimation uncertainty in model parameters is difficult \todo{cite, elaborate}.

\subsection{Summary}

Our simulation results confirm the theoretical guarantee of KALE dominating KILE in prediction MSE when the covariance function is known, and furthermore HMC dominating KALE. The magnitude of differences in MSE between these methods is greatest when the process is moderately smooth relative to the spatial sampling (\textit{e.g.}, $0.01 \leq \beta \leq 0.5$), when the magnitude of location errors $\sigma^2_u$ is largest, and when nugget variation $\sigma^2_x$ is smallest. For such regions of the parameter space, KILE fails to deliver prediction intervals with proper coverage, whereas KALE and HMC can give valid prediction intervals for any parameter values.

An important consequence in adjusting for location errors with a known covariance function is the corresponding adjustment to the nugget. The discussion in (Sections 3.6 and 3.7 of) \citet{stein1999interpolation}  emphasizes the importance of correctly specifying the high-frequency behavior of the process when interpolating (correctly specifying the low-frequency behavior is less crucial), including the nugget term. Estimating parameters, including the nugget term $\sigma^2_x$, implicitly corrects for model misspecification when ignoring location errors. Thus we see little difference in predictive performance between KALE and KILE when parameters are first estimated. Depending on the choice of prior, KALE/KILE may give lower MSE predictions than HMC, which averages over posterior parameter uncertainty; however, interval coverage is better for HMC (using weak prior information) than for KALE/KILE.

\section{Interpolating Northern Hemisphere Temperature Anomolies}
\label{sec:cru}

To illustrate the methods discussed in this paper, we consider interpolating northern hemisphere temperature anomolies during the summer of 2011 using the publicly available CRUTEM3v data set\footnote{http://www.cru.uea.ac.uk/cru/data/temperature/} [\citet{brohan2006uncertainty}]. Figure \ref{fig:cru_data_plot} shows our data. These data are used in geostatistical reconstructions of the Earth's temperature field, which interpolate temperatures at unobserved points in space-time in order to better understand the historical behavior of climate change (see, \textit{e.g.}, \citet{tingley2010bayesian} and \citet{richard2012new}). Each observation is a spatiotemporal average: temperature readings are averaged over the April--September period and each $5^{\circ} \times 5^{\circ}$ longitude-latitude grid cell. These values are then expressed as anomolies relative to the global average during the period 1850--2009, which is calculated using an ANOVA model [\citet{tingley2012bayesian}]. Apart from this spatiotemporal averaging, numerous other preprocessing steps adjust this data for differences in altitude, timing, equipment, and measurement practices between sites, along with other potential sources of error; please see \citet{morice2012quantifying} and \citet{jones2012hemispheric} for more details. 

Our analysis, restricted to interpolating a single year of data, and without using external data such as temperature proxies [\citet{mann2008proxy}], is intended as a proof of concept rather than as a refinement or improvement to existing analyses of these data. We wish to illustrate the potential impact of location errors on conclusions drawn from these data.

\begin{knitrout}
\definecolor{shadecolor}{rgb}{0.969, 0.969, 0.969}\color{fgcolor}\begin{figure}[H]

\includegraphics[width=\maxwidth]{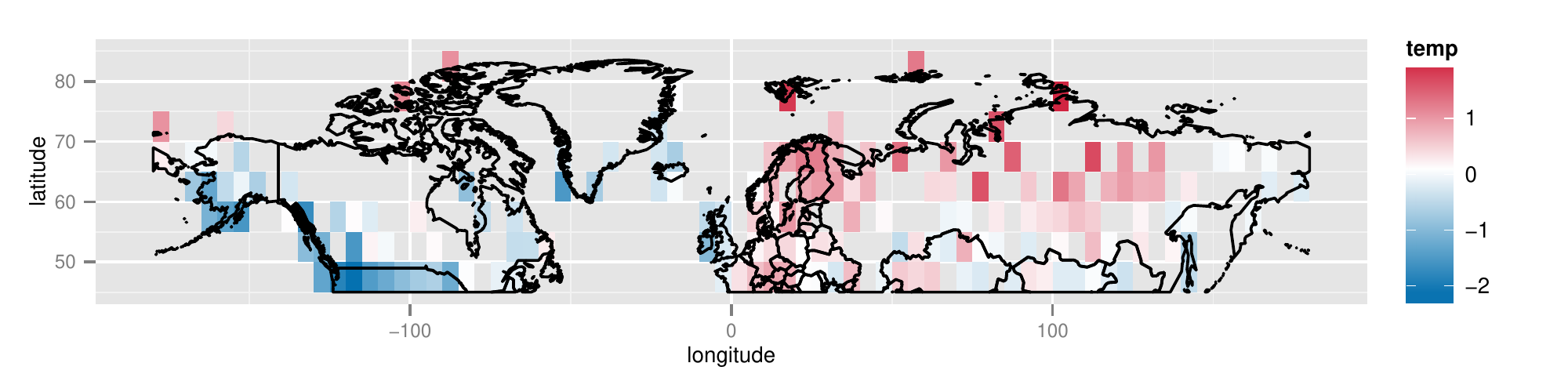} \caption[CRUTEM3v data for summer 2011, with 2011 mean subtracted so that measurements represent spatial anomolies]{CRUTEM3v data for summer 2011, with 2011 mean subtracted so that measurements represent spatial anomolies. Generally speaking, we see lower (cooler) anomolies in North America and positive (warmer) anomolies in Europe. Higher latitudes also tend to have positive anomolies.\label{fig:cru_data_plot}}
\end{figure}

\end{knitrout}

% figure of data

The ``gridding'', or spatial averaging across $5^{\circ} \times 5^{\circ}$ cells, complicates analyses using Gaussian process models [\citet{director2015connecting}]. However, assuming a smooth temperature field, we know that the recorded spatial average must be realized exactly at some location in each grid box (closer to the center if a lot of points have been averaged together). This frames the spatial averaging problem as a location measurement error problem: instead of observing the temperature $x(s)$ at each grid center $s$, we observe the temperature at an unknown location displaced from the grid center: $y(s) = x(s + u)$.

Following \citet{tingley2010bayesian}, we assume an exponential covariance function for $x(s)$, where distance is calculated along the Earth's surface. As $s$ is given in terms of longitude/latitude ($s = (\psi, \phi)$), this has the form
\begin{align}
c(s_1, s_2) &= \tau^2 \exp(-\beta \Delta) + \sigma^2_x \mathbf{1}_{s_1 = s_2}\nonumber \\
\Delta & = 2r\arcsin \sqrt{\sin^2\left(\frac{\phi_2 - \phi_1}{2}\right) + \cos(\phi_1) \cos(\phi_2)\sin^2\left(\frac{\psi_2 - \psi_1}{2}\right)}, \label{haversine}
\end{align}
where $r=6371$ is the radius of the earth (in km). At higher latitudes ($\phi$), the centers of each grid cell are closer together, so nearby observations are more strongly correlated. The nugget term $\sigma^2_x$ represents some combination of measurement error in temperature readings and high-frequency spatial variation that is inestimable using the gridded observation samples. 

We assume the following model for location errors $u_i$, which are additive displacements of longitude/latitude coordinates $s_i = (\psi_i, \phi_i)$:
\begin{equation}\label{u_geo}
u_i \sim \N \left( \mathbf{0}, \sigma^2_u \left(\frac{180}{\pi r}\right)^2 \begin{pmatrix}\frac{1}{\cos^2(\phi_i)} & 0 \\ 0 & 1 \end{pmatrix} \right).
\end{equation}
This prior is equivalent to assuming that distance along the Earth's surface (great-circle distance) between each grid center and the corresponding observation location has a scaled chi distribution, $d(s_i, s_i + u_i) \sim \sigma_u \chi$. Combining \eqref{u_geo} and \eqref{haversine}, we use Monte Carlo to compute $k$.

We treat parameters $\tau^2, \beta, \sigma^2_x$ as unknown, but fix $\sigma^2_u = 7500$. At this value, the median magnitude of the location errors in great-circle distance is 102km, which is consistent with analyzing the coordinates of the temperature recording sites used to compile the CRUTEM3v data\footnote{Station locations are vieweable at https://www.ncdc.noaa.gov/oa/climate/ghcn-daily/}. %Also, with this choice of $\sigma^2_u$, the chance that $s_i + u_i$ is within the innermost 20\% region of the corresponding grid rectangle is between 90\% (for small area rectangles near the north pole) and 100\% (for lower latitudes).

\subsection{Kriging}
We first apply Kriging approaches to interpolate the CRUTEM3v data, both adjusting for and ignoring location errors \eqref{u_geo}. Because parameters $\tau^2, \beta, \sigma_x^2$ are unknown, we first need to estimate them using maximum likelihood (when ignoring location errors) or maximum pseudo-likelihood \eqref{pseudo} (when adjusting for location errors). These can then be plugged in to covariance functions $c$ and $k$ to obtain ``empirical'' Kriging equations we can use for interpolation [\citet{zimmerman1992mean}]. 

\begin{figure}[h!]
\centering
   \includegraphics[width=1.0\textwidth]{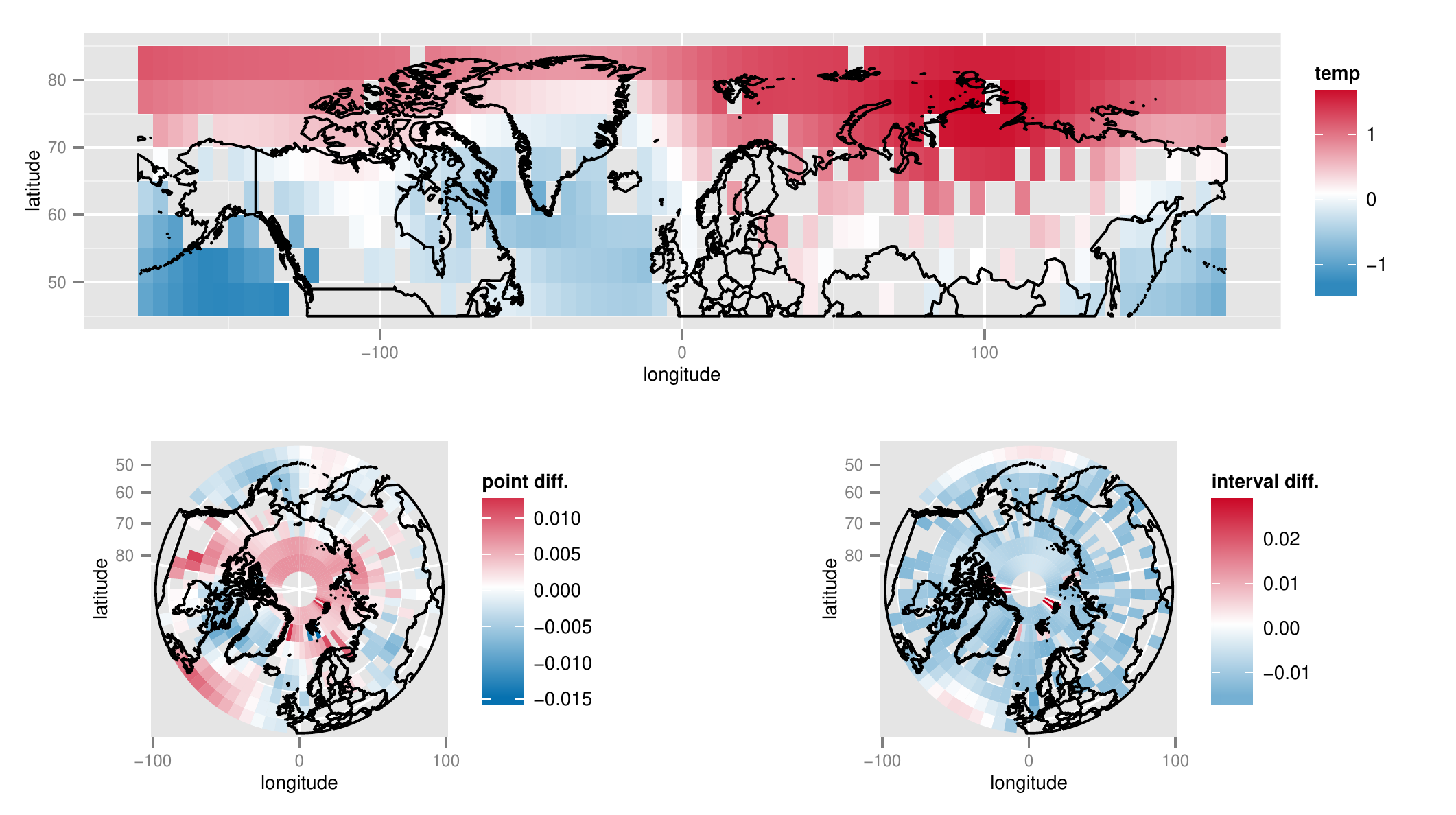}
\caption[ Kriging interpolations for CRUTEM3v data]{Kriging results for interpolating temperature anomolies from summer 2011. The top plot shows interpolations at unobserved grid centers given by \emph{KALE}. The bottom left plot shows the difference in estimates between \emph{KALE} and \emph{KILE} (\emph{KALE} -- \emph{KILE}), and the bottom right plot shows difference in 95\% interval widths between \emph{KALE} and \emph{KILE}.} \label{fig:cru_krig}
\end{figure}

We find small differences in parameter estimates when ignoring location errors (assuming $\sigma_u^2 = 0$) and adjusting for them (assuming $\sigma_u^2 = 7500$):
\begin{table}[h!]
\begin{center}
\begin{tabular}{rrrr}
\toprule
$\sigma_u^2$ & $\hat{\tau}^2$ & $\hat{\beta}$ & $\hat{\sigma}_x^2$ \\
\midrule
0 & 1.1671 & \ensuremath{1.4275\times 10^{-4}} & 0.0747 \\
7500 & 1.1649 & \ensuremath{1.4677\times 10^{-4}} & 0.0699 \\
\bottomrule
\end{tabular}
\end{center}
\caption[ Covariance function parameter estimates using Kriging]{Covariance function parameter estimates when ignoring location errors (assuming $\sigma_u^2 = 0$) and adjusting for location errors (assuming $\sigma_u^2 = 7500$).}
\label{paramtable}
\end{table}
%\todo{provide SEs from information matrix}
%This suggests that location-error induced covariance function $k$ is very close to $c$ for all pairs of sample locations---this is expected when the magnitude of location errors is small relative to the spatial sampling scheme. Thus, the likelihood and pseudo-likelihood provide nearly the same parameter estimates. 
\newline
Consequently, when we interpolate data at the centers of grid cells for which no data was observed, we see differences between the KALE and KILE approaches. Figure \ref{fig:cru_krig} shows the differences between KALE and KILE interpolations (both point and interval estimates). Relative to the range of the data (most anomolies are in the interval $(-1, 1)$), the discrepency between KALE and KILE does not seem very significant.

\subsection{HMC}

\begin{figure}[h!]
\centering
   \includegraphics[width=1.0\textwidth]{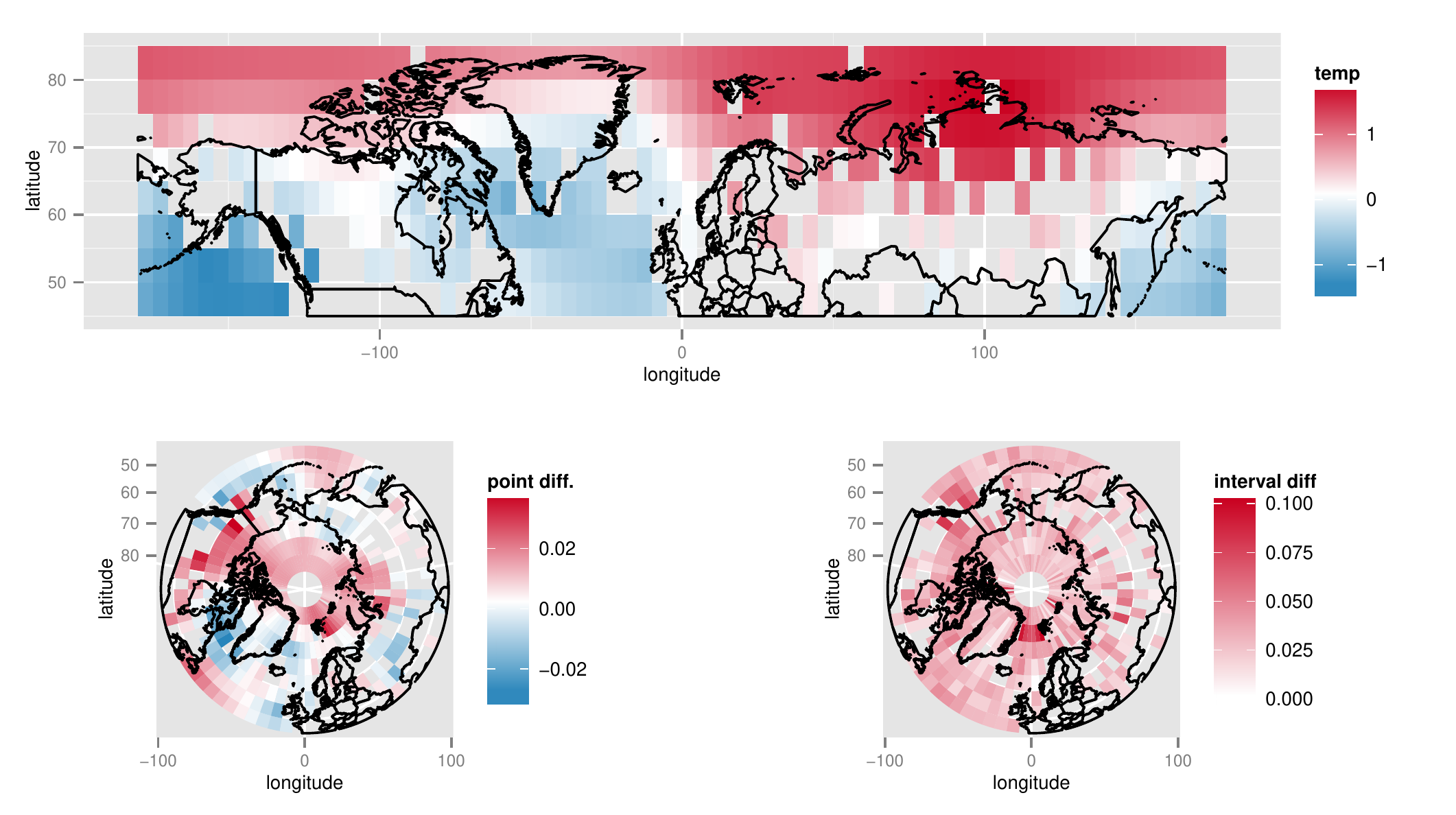}
\caption[ HMC interpolations CRUTEM3v data]{Results for interpolating temperature anomolies from summer 2011 using HMC. The top plot shows interpolations at unobserved grid centers, assuming location errors $\sigma^2_u = 7500$. The bottom left plot shows the difference in estimates between the location error model and the model with $\sigma^2_u = 0$. The bottom right plot shows difference in 95\% interval widths.} \label{fig:cru_hmc}
\end{figure}

Using HMC, parameter inference and interpolations are made simultaneously. The resulting point and interval predictions differ substantially from the Kriging results. However, because HMC incorporates parameter uncertainty in predictions, this comparison is not sufficient to illustrate the impact of location errors on conclusions from this data. A more appropriate comparison is between HMC with a location error model ($\sigma_u^2 = 7500$), and HMC assuming with no location errors ($\sigma_u^2 = 0$). These results are plotted in Figure \ref{fig:cru_hmc}. 

Using HMC, accounting for location errors produces more significant differences in inference/prediction than was observed for Kriging. This is particularly true for interval predictions, where adjusting for location errors yields intervals as much as 0.1 wider, which is a significant discrepency when most observations lie in $(-1, 1)$. 
%Unlike the results comparing KALE and KILE in Figure \ref{fig:cru_krig}, point estimates and interval estimates are dramatically impacted by adjusting for location errors ($\sigma^2_u = 7500$ vs. $\sigma^2_u = 0$). Point estimates differ by as much as $0.5$ degrees, which is significant when the entire data set spans $(-2, 2)$ degrees. Similarly, $95\%$ interval widths differ by as much as $0.75$ degrees between the $\sigma^2_u = 500$ and $\sigma^2_u = 0$ models.

\begin{figure}[h!]
\centering
   \includegraphics[width=1.0\textwidth]{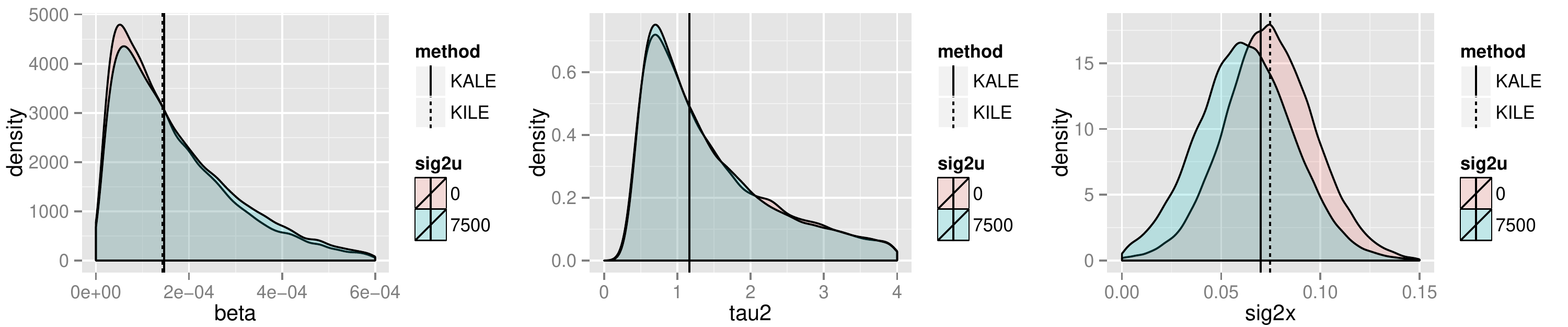}
\caption[ Posterior density for covariance parameters of CRUTEM3v data]{Density of posterior draws from HMC using $\sigma^2_u = 7500$ (blue) and $\sigma^2_u = 0$ (red). Point estimates of these parameters from Kriging (Table \ref{paramtable}) are shown as vertical lines.} \label{fig:cru_params}
\end{figure}

Figure \ref{fig:cru_params} shows posterior densities for unknown parameters of the covariance function based on HMC draws from the $\sigma^2_u = 7500$ and $\sigma^2_u = 0$ models (the Kriging estimates of these parameters are vertical lines). HMC under location error model ($\sigma^2_u = 7500$) gives slightly larger $\beta$ estimates than when using $\sigma^2_u = 0$, meaning observations are inferred to be less strongly correlated. This yields prediction intervals that tend to be wider (see Figure \ref{fig:cru_hmc}). The most extreme descrepencies occur in the arctic, where distances between grid points are closest. The fact that modeling location errors adds additional uncertainty to arctic predictions is of particular interest to climate scientists, as accurate climate reconstruction for the arctic region is essential for understanding recent climate change patterns [\citet{cowtan2014coverage}].

The difference between predictions obtained under the $\sigma^2 = 0$ and $\sigma^2_u = 7500$ models using HMC suggests that modeling location errors, even when they are small in magnitude, meaningfully impacts parameter estimates and predictions at unobserved locations. The fact that results for HMC (assuming $\sigma^2_u = 7500$) also differ from the results using KALE, while the KILE results do so less, demonstrates that moment procedures such as Kriging may be ineffective in adjusting for these errors.

% analysis using HMC: sig2u = 500 vs sig2u = 0, mean and interval width and parameter estimates
% kriging: mean and interval width
%\section{Methods for Larger Data Sets}
\section{Conclusion}
\label{sec:conclusion}
In this paper, we have explored the issue of Gaussian process regression when locations in the input space $\Ss$ are subject to error. Even when location errors are quite small in magnitude, it is essential to adjust Kriging equations in order to obtain good point and interval estimates; further improvements can be made by using MCMC to sample directly from the distribution of the measurement of interest given the sampled data. 

%While modeling location errors is important, it is difficult to recognize their presence from the data alone. Parameters of the location error distribution are not identifiable from the parameters of the covariance function, so they must be inferred separately or fixed at reasonable values.

Both MCMC and Kriging will be infeasible for large data sets, due to the cost of the covariance matrix inversion. A useful future study would be to adapt the procedures discussed in this paper to methods for inference and prediction for large spatial data sets, such as the predictive process approach [\cite{banerjee2008gaussian}], low rank representations [\citet{cressie2008fixed}], likelihood approximations [\citet{stein2004approximating}], and Markov random field approximations [\citet{lindgren2011explicit}]. It will also be useful to extend the analysis of this paper to regimes where location errors may be correlated with the process of interest $x$. For example, in climate data, regions with extreme climates will be harder to sample, thus there may be greater error in the spatial refencing of such sampling than for regions that are easier to sample.

\appendix
\section{Proofs of results}

\subsection{Proof of Proposition \ref{validcovfn}}
$k$ is a valid covariance function if and only if for all $n$, $\bs_n$, and $\{a_i \in \R, i=1, \ldots, n\}$, we have 
$$\sum_{i=1}^n \sum_{j=1}^n a_i a_j k(s_i, s_j) \geq 0.$$ 
From \eqref{cov_y}, this condition can be rewritten:
\begin{align*}
\sum_{i=1}^n \sum_{j=1}^n a_i a_j k(s_i, s_j) &= \sum_{i=1}^n \sum_{j=1}^n a_i a_j \int_{\Ss} c(s_i + u_i, s_j + u_j) dg_{\bs_n}(\bu_n) \\
 &= \int_{\Ss} \sum_{i=1}^n \sum_{j=1}^n a_i a_j c(s_i + u_i, s_j + u_j) dg_{\bs_n}(\bu_n)
\end{align*}
As $c$ is a valid covariance function, the integrand in this expression is always non-negative, so the integral is also non-negative. Thus $k$ is a valid covariance function.

Note that for the common scenario where location errors are independent, so that $g_{\bs_n}$ is a product measure ${g_{s_1} \times \ldots \times g_{s_n}}$, then Proposition \ref{validcovfn} is a special case of kernel convolution [\citet{rasmussen2006gaussian}].

\subsection{Proof of Proposition \ref{prop1}} 
Without loss of generality, we can assume $\tau^2 = 1$ and fix $\beta, \Delta >0$. Using the fact that $k^*(s, s^*) = \E[\exp(-\beta \|s + u - s^*\|^2)]$, evaluating the moment generating function of a non-central $\chi_p^2$ random variable $\|s + u - s^*\|^2$ yields
$$c(\sigma_u^2) \equiv \E[(\xkale(s^*) - x(s^*))^2] = 1 - \left(\frac{1}{1 + 2\beta\sigma^2_u}\right)^p \exp \left( \frac{-2\beta\Delta^2}{1 + 2\beta\sigma_u^2} \right).$$
Differentiating, we get
$$c'(\sigma_u^2) = \frac{2\beta [2\beta(p\sigma^2_u - \Delta^2) + p]}{(1 + 2\beta\sigma_u^2)^{p + 2}} \exp \left( \frac{-2\beta \Delta^2}{1 + 2\beta\sigma_u^2}\right).$$
If $\beta \Delta^2 \leq p/2$, then $c'(\sigma_u^2) > 0$ for all $\sigma_u^2 > 0$. Since $c(\sigma_u^2)$ is left continuous at 0, continuous on $\R_+$, and $c(0) = c_0$, this means $\beta \Delta^2 \leq p/2$ implies $c(\sigma_u^2) \geq c_0$ for all $\sigma_u^2$.

Otherwise, if $\beta \Delta^2 > p/2$, then for all $0 < \sigma^2_u < \frac{\Delta^2}{k} - \frac{1}{2\beta}$, we have $c'(\sigma^2_u) < 0$. Once again, because $c(\sigma_u^2)$ is left continuous at 0, continuous on $\R_+$, and $c(0) = c_0$, this means $c(\sigma_u^2) < c_0$ for $\sigma_u^2$ in this interval.

\subsection{Proof of Proposition \ref{coverprop}}
Let $W = x(s^*) - \xkale(s^*)$. We can explicitly write the dependence of $W$ on $\bu_n$:
\begin{equation*} %\label{Wdist}
W | \bu_n \sim \N ( 0, V(\bu_n))
\end{equation*}
where
\begin{align*}
V(\bu_n) & = \sigma^2 + \gamma'\C(\bs_n + \bu_n, \bs_n + \bu_n)\gamma - 2\gamma'\C(\bs_n + \bu_n, s^*), \\
\gamma &= \K(\bs_n, \bs_n)^{-1}\K^*(\bs_n, s^*),
\end{align*}
and $\sigma^2 = \V [x(s^*)]$. Thus 
\begin{align*}
\prob(W < z) &= \E[\prob(W < z | \bu_n)] \\
 &=  \E \left[ \Phi \left( \frac{z}{\sqrt{V(\bu_n)}} \right) \right].
\end{align*}

\subsection{Proof of Theorem \ref{theorem1}}
First, our assumptions in the hypothesis imply that $k$ is continuous everywhere in $\Ss^2$ except where $s_1 = s_2$. To see this, take any distinct $s_1, s_2 \in \Ss$ and sequence $s_1^m, s_2^m$ converging to $(s_1, s_2)$. The sequence $c(s_1^m + u_1^m, s_2^m + u_2^m)$ is bounded and converges in distribution to $c(s_1 + u_1, s_2 + u_2)$. Thus, by the Dominated Convergence theorem, $k(s_1^m, s_2^m) \rightarrow k(s_1, s_2)$. 

Now, for any $n \in \mathbb{N}$, the KILE MSE in predicting $x(s^*)$ given $\by_n$ is 
\begin{align}
\E[(x(s^*) - \xkile(x^*))^2] &= \V[x(s^*)] - 2\C(s^*, \bs_n) \C(\bs_n, \bs_n)^{-1} \K^*(\bs_n, s^*) \nonumber \\
 & \hspace{-1cm} + \C(s^*, \bs_n) \C(\bs_n, \bs_n)^{-1}\K(\bs_n, \bs_n) \C(\bs_n, \bs_n)^{-1} \C(\bs_n, s^*).
 \label{kile_mse}
 \end{align}
%{\tcr{typo above? v?}} {\dc{Changed to $\V$ for variance}} 
The matrix $\C(\bs_n, \bs_n)$ is symmetric and positive definite, and thus it can be written as $\C(\bs_n, \bs_n) = \bQ \bLam \bQ'$, where $\bQ$ is an orthogonal matrix and $\bLam$ is diagonal. Assume without loss of generality the entries are $\bLam$ are ordered $0 < \lambda_1 < \ldots < \lambda_n$. Similarly, write $\K(\bs_n, \bs_n) = \bR \bOm \bR'$. Further letting $\ba = \C(s^*, \bs_n) \bQ$, $\bb = \bQ' \K(\bs_n, s^*)$, and $\bD = \bQ'\bR \bOm^{\frac{1}{2}}$, we can write \eqref{kile_mse} as
\begin{align}
\E[(x(s^*) - \xkile(x^*))^2] &= \V[x(s^*)] - 2\ba' \bLam^{-1} \bb + \ba' \bLam^{-1} \bD \bD' \bLam^{-1} \ba \nonumber \\
&= \V[x(s^*)] - 2\sum_{i=1}^n \frac{a_ib_i}{\lambda_i} + \sum_{j=1}^n \left( \sum_{i=1}^n \frac{a_i D_{ij}}{\lambda_i} \right)^2.
\label{kile_mse_2}
\end{align}
Let $\xi = \lambda_1^{-1}$; then Equation $\eqref{kile_mse_2}$ can be expressed as 
%{\tcr{here, $a_{(1)}$ and $D_{(1), j}$ are not defined!}} {\dc{Fixed: we can just assume WLOG $\lambda_1$ is smallest eigenvalue}}
\begin{equation}
\E[(x(s^*) - \xkile(x^*))^2] = \xi^2 \left(\sum_{j=1}^n a_1^2 D_{1 j}^2\right) + h(\xi),
\label{quadratic}
\end{equation}
where $h$ is linear in $\xi$.
%{\tcr{we need $h$ to be positive?}} {\dc{No, since it is always dominated by the quadratic term $\xi^2$.}}

Without loss of generality, assume $s_1 \rightarrow s_2$. Thus $\C(\bs_n, \bs_n)$ becomes rank $n-1$, with $\lambda_{1} \rightarrow 0$ and for all $i > 1$, $\lambda_{i} \rightarrow \lambda^*_{i} > 0$. Thus $\xi \rightarrow \infty$. However, $\K(\bs_n, \bs_n)$ does not become singular, since $\prob(u_1 \neq u_2) < 1$ implies 
$$\lim_{s_1 \to s_2} k(s_1, s_2) \neq \V[y(s_2)].$$
%$$\lim_{s_1 \rightarrow s_2} k(s_1, s_2) = \lim_{s_1 \rightarrow s_2} \E[c(s_1 + u_1, s_2 + u_2)] = \E[c(s_2 + u_1, s_1 + u_2)] \neq \V[y(s_2)]. $$
Since $k$ is continuous and $\K(\bs_n, \bs_n)$ nonsingular in the limit, all of the terms besides $\xi$ in \eqref{kile_mse_2} converge as $s_1 \rightarrow s_2$; that is $\ba \rightarrow \ba^*$, $\bb \rightarrow \bb^*$, and $\bD \rightarrow \bD^*$ as $s_1 \rightarrow s_2$. Moreover, we cannot have $D^*_{1 j} = 0$ for all $j$, as this contradicts $\K(\bs_n, \bs_n)$ remaining full-rank. Lastly, since $\C(s^*, \bs_n) \neq \mathbf{0}$ and $\bQ$ is orthogonal, $a_i \neq 0$ and $a^*_i \neq 0$ for all $i=1, \ldots, n$.

Thus the quadratic coefficient in \eqref{quadratic}, $\sum_{j=1}^n a_1^2 D_{1 j}^2$ is strictly positive, and $h(\xi) = \mathcal{O}(\xi)$. Because $\xi \rightarrow \infty$, we get
$$\lim_{s_1 \to s_2} \E[(x(s^*) - \xkile(x^*))^2] = \infty.$$

For pathological choices of $g_{\bs_n}$ where $k$ is not continuous everywhere and limits for $\bb$ and $\bD$ may not exist, all components of these terms can be still be bounded, which is sufficient for Theorem \ref{theorem1} to hold.

\subsection{Proof of Proposition \ref{morris_prop}}

Bayes rule predictors by definition satisfy $\mathrm{R}_{\pi}(\pi) \leq \mathrm{R}_{\pi}(\tilde{\pi})$, which confirms the two inequalities in the statement of Proposition \ref{morris_prop}. The equality $\mathrm{R}_{\pi}(\pi_0) = \mathrm{R}_{\pi_0}(\pi_0)$ holds since the risk of the Bayes estimator under $\pi_0$ is a quadratic form, and therefore constant for all $\pi \in \Pi_{\mathbf{0}, \C}$:
\begin{align*}
\mathrm{R}_{\pi_0}(\pi_0) &= \E_{\pi}[(\E_{\pi_0}[x(s^*)|\bx_n ] - x(s^*))^2] \\
 &= \E_{\pi}[(\C(s^*, \bs_n)\C(\bs_n, \bs_n)^{-1}\bx_n - x(s^*))^2] \\
 &= c(s^*, s^*) - \C(s^*, \bs_n)\C(\bs_n, \bs_n)^{-1}\C(\bs_n, s^*) \\
 &= \mathrm{R}_{\pi}(\pi_0).
 \end{align*}

\section*{Acknowledgements}{We thank Luke Bornn and Peter Huybers for helpful comments and
encouragement. NSP was partially supported by an ONR grant. DC was partially supported by a research grant from the Harvard University Center for the Environment.}
\bibliographystyle{imsart-nameyear}

\bibliography{gpreg_bib}

\end{document}